\documentclass[journal]{IEEEtran}

\usepackage[T1]{fontenc}
\usepackage{graphicx}
\usepackage{cite}
\usepackage{subcaption}
\usepackage{overpic}
\usepackage{amssymb}
\usepackage[cmex10]{amsmath}
\usepackage{amsthm}
\usepackage{todonotes}
\usepackage{stfloats}
\usepackage{bbm}
\usepackage{color}
\usepackage{lipsum}

\usepackage{mathtools}

\DeclarePairedDelimiter\floor{\lfloor}{\rfloor}

\IEEEoverridecommandlockouts

\def\rank{\mathrm{rank}}
\def\sinc{\mathrm{sinc}}
\def\Htran{\mbox{\tiny $\mathrm{H}$}}
\def\Ttran{\mbox{\tiny $\mathrm{T}$}}
\def\imagunit{\mathsf{j}} 

\theoremstyle{plain}

\newtheorem{corollary}{Corollary}

\newtheorem{remark}{Remark}
\newtheorem{assumption}{Assumption}

\newcommand{\vect}[1]{{\bf{#1}}}

\begin{document}

\title{\huge{Wavenumber-Division Multiplexing in Line-of-Sight Holographic MIMO Communications}}

\author{
\IEEEauthorblockN{Luca Sanguinetti, \emph{Senior Member, IEEE}, Antonio Alberto D'Amico, Merouane Debbah, \emph{Fellow, IEEE}
\thanks{
\newline \indent L.~Sanguinetti and A. A. D'Amico are with the Dipartimento di Ingegneria dell'Informazione, University of Pisa, Pisa, Italy (email: luca.sanguinetti@unipi.it, antonio.damico@unipi.it).
\newline \indent M. Debbah is with the Technology Innovation Institute, 9639 Masdar City, Abu Dhabi, United Arab Emirates (email: merouane.debbah@tii.ae) and also with CentraleSupelec, University Paris-Saclay, 91192 Gif-sur-Yvette, France.}
}}
\maketitle

\begin{abstract}
Starting from first principles of wave propagation, we consider a multiple-input multiple-output (MIMO) representation of a communication system between two spatially-continuous volumes. This is the concept of holographic MIMO communications. The analysis takes into account the electromagnetic interference, generated by external sources, and the constraint on the physical radiated power. 
The  electromagnetic MIMO model is particularized for a pair of parallel line segments in line-of-sight conditions.
Inspired by orthogonal-frequency division-multiplexing, we assume that the spatially-continuous transmit currents and received fields are represented using the Fourier basis functions. In doing so, a wavenumber-division multiplexing (WDM) scheme is obtained, which is not optimal but can be efficiently implemented. The interplay among the different system parameters (e.g., transmission range, wavelength, and sizes of source and receiver) in terms of number of communication modes and level of interference among them is studied with conventional tools of linear systems theory. Due to the non-finite support (in the spatial domain) of the electromagnetic channel, WDM cannot provide non-interfering communication modes. 
The interference decreases as the receiver size grows, and goes to zero only asymptotically. Different digital processing architectures, operating in the wavenumber domain, are thus used to deal with the interference. 
The simplest implementation provides the same spectral efficiency of a singular-value decomposition architecture with water-filling when the receiver size is comparable to the transmission range. The developed framework is also used to represent a communication scheme that performs only an integration over short spatial segments. This is equivalent to a classical MIMO system with uniform linear arrays made of electrically small dipoles. Numerical comparisons show that better performance than WDM can be achieved only when a higher number of radio-frequency chains is used.
\end{abstract}


\section{Introduction}

In the quest for ever-increasing data rates, it is natural to continue looking for more bandwidth, which in turn pushes the operation towards higher frequencies than in the past. The current frontier is in bands up to 86\,GHz. However, there is at least 50\,GHz of suitable spectrum in the range 90--200\,GHz and another 100\,GHz in the range 220--320\,GHz. {These bands are commonly referred to as sub-THz bands and are currently receiving attention for a large variety of emerging applications 
both of short-range nature (inter-chip communications, data center interconnections, indoor local area networks) and also of longer range (wireless backhaul, satellite interconnection), e.g.~\cite{Rappaport2019,Heedong_Com_Mag}.}
As we move up towards high frequencies, the transmission range necessarily shrinks and line-of-sight (LoS) propagation becomes predominant. Classical results in the multiple-input multiple-output (MIMO) literature teach that spatial multiplexing is inevitably compromised in LoS propagation~\cite{TseBook}. To understand why this is not the case, recall that such {classical results rely on the planar wavefront approximation over the array~\cite{Foschini-1999}; that is, all the antenna elements receive the signal from the same angle of arrival with the same  propagation path loss.} This approximation becomes inadequate at high frequencies. Since the wavelength reduces dramatically and the transmission range tends to be short, the wave curvature over the array is no longer negligible and the LoS MIMO channel matrix becomes of high rank~\cite{Foschini-1999,Jeng-Shiann-2005,Bohagen2007,Bohagen-2009,Madhow-2011,Heedong-2021}.  By properly modeling the wave curvature over the array, one  can compute the effective number $\eta$ of degrees-of-freedom (DoF), which represents the number of independent data streams that can be reliably sent over the channel in the high signal-to-noise regime. For the two parallel line segments illustrated in Fig.~\ref{figure_linear_arrays}, if $L_s, L_r \ll d$ we have that (e.g.,~\cite{Miller:00})
\begin{align}\label{eq:paraxial_approximation}
\eta = \frac{L_s L_r}{\lambda d}
\end{align}
which depends on the solid angles $\Omega_r=L_r/d$ and $\Omega_s = L_s/d$. This result is obtained from the paraxial approximation~\cite{Miller:00}, which follows from elementary diffraction theory. 

%
\subsection{Motivation and Contribution}
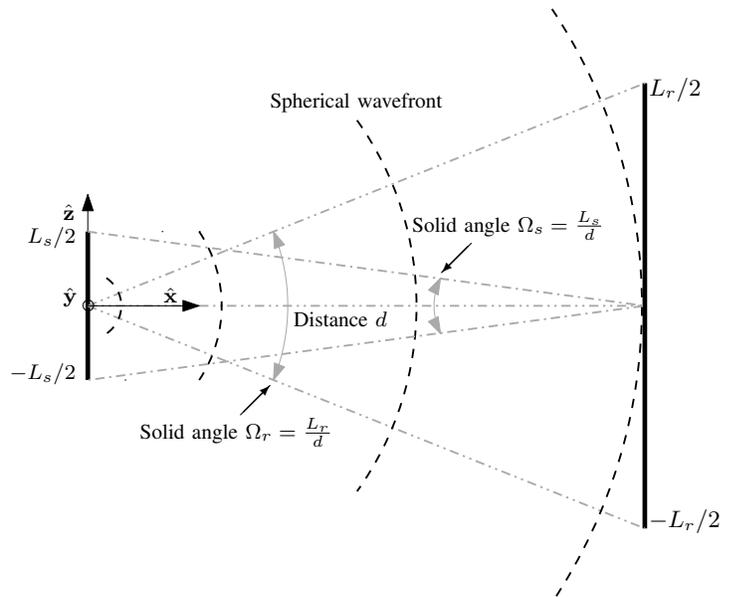
\begin{figure}[t!]
	\centering 
	\begin{overpic}[width=0.85\columnwidth,tics=10]{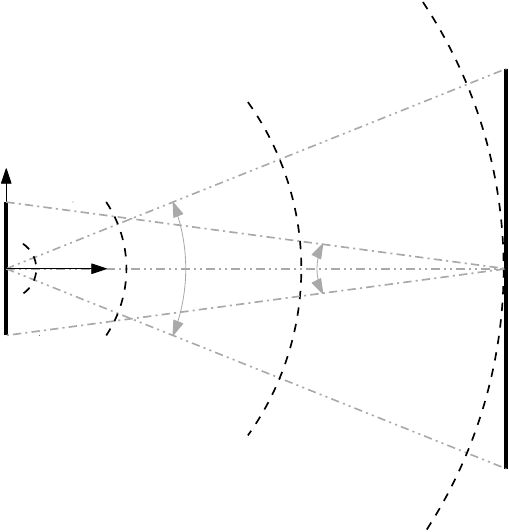}
	\put(10,27){\footnotesize Solid angle $\Omega_r=\frac{L_r}{d}$}
        \put(56,62){\footnotesize Solid angle $\Omega_s=\frac{L_s}{d}$}
        \put(66,60){\vector(-1, -1){5}}
	\put(32,83){\footnotesize Spherical wavefront}
	\put(27,31){\vector(1, 1){5}}
	\put(96,85){\small $L_r/2$}
	\put(96,12){\small $-L_r/2$}
	\put(14,50){\footnotesize $\hat{\bf x}$}
	\put(-3,50){\footnotesize $\hat{\bf y}$}
	\put(-3,64){\footnotesize $\hat{\bf z}$}
	\put(1.3,49.5){\circle{2}}
	\put(36,46){\footnotesize Distance $d$}
	\put(-9,60){\footnotesize $L_s/2$}
	\put(-12,37){\footnotesize $-L_s/2$}
\end{overpic}
	\caption{Illustration of the approach to estimating the number of degrees-of-freedom in the communication between two linear segments of size $L_s$ and $L_r$, at a distance $d$. The solid angle $\Omega_r$ subtended by the receive line segment of length $L_r$ at a perpendicular distance $d$ is $\Omega_r={L_r}/{d}$. The solid angle $\Omega_s$ is similarly given by $\Omega_s={L_s}/{d}$.}
	\label{figure_linear_arrays} 
\end{figure}

Motivated by the need of capturing the fundamental properties of wave propagation in wireless communications, we introduce the MIMO representation of a communication system between two \emph{spatially-continuous} volumes of arbitrary shape and position. The model is developed on the basis of prior analyses in electromagnetic theory (e.g., \cite{Miller:00,Marengo2008,Wallace2008,franceschetti_2017}). Particularly, starting from the Maxwell's equations describing electromagnetic fields in any known (i.e., deterministic) channel medium, the MIMO model is obtained by representing the spatially-continuous transmit currents and received fields with arbitrary sets of basis functions. {This is the concept of \emph{holographic} communications~\cite{Sanguinetti2021,DardariHolographic,Huang2020}, known also in the literature as continuous-aperture MIMO (CAP-MIMO) communications\cite{Sayeed2010,Sayeed2011}, or large intelligent surface (LIS) communications~\cite{Edfors2018}}; that is, the capability to generate any current density distribution (in order to obtain the maximum flexibility in the design of the radiated electromagnetic field) and to weight the impinging electric field according to a desired function. The analysis takes into account the {electromagnetic interference (EMI)} generated by external sources~\cite{Wallace2008}, and emphasizes the relation between the energy of transmit currents and the physical radiated power. The ultimate performance limit of such MIMO model with spatially-uncorrelated Gaussian noise is known~\cite{Miller:00,Marengo2008,Wallace2008}, and can be achieved by using the eigenfunctions of channel operators as basis sets. In analogy to classical analysis of time- and band-limited systems, the optimal eigenfunctions take the form of prolate spheroidal wave functions~\cite{Slepian64}, whose computation and implementation are in general prohibitive~\cite{Dardari-2020}. {Our contribution here is thus not to look at the fundamental limits, which are known (under the aforementioned conditions), but to provide a unified representation of the electromagnetic wave communication problem, which can be directly used by communication theorists to design and study implementable (from a technological point of view) communication schemes. This is exactly at the basis of the major contribution of this paper, which consists in particularizing the developed model for the LoS system in Fig.~\ref{figure_linear_arrays} and in proposing a wavenumber-division multiplexing (WDM) scheme, inspired by orthogonal-frequency division-multiplexing (OFDM).} This is achieved by assuming that the spatially-continuous transmit currents and received fields are represented using the Fourier basis functions, {which can be efficiently implemented at the electromagnetic level by means of lens antenna arrays or metasurfaces (e.g.,~\cite{Tretyakov2015,Li_2015,Zhang2016a,Chan-Byoung2018})}. {This is also the concept of beamspace MIMO~\cite{Sayeed2013a}, which transforms the signals in antenna space to beamspace and exploits its lower dimension to significantly reduce the number of radio-frequency (RF) chains required.}
The properties of WDM (in terms of number of communication modes and level of interference) and the functional dependence on the different system parameters (e.g., transmission range, wavelength, and sizes of source and receiver) are studied using the conventional tools of linear systems theory. Unlike OFDM, we show that WDM cannot convert the MIMO channel into a set of parallel and independent sub-channels due to the non-finite support (in spatial domain) of the electromagnetic
channel response. The orthogonality among the communication modes is achieved only when the size of the receiver grows infinitely large. {Different digital processing architectures, operating in the wavenumber domain, are thus used to deal with the interference. We show that the simplest implementation, represented by a bank of one-tap complex-valued multipliers, provides the same spectral efficiency (SE) of a singular-value decomposition (SVD) architecture with water-filling when the receiver size is comparable to the transmission range. Comparisons are also made between WDM and {a MIMO system that performs only an integration of the electric field over short spatial segments. This is equivalent to a MIMO system equipped with uniform linear arrays made of (equally spaced) elementary dipoles of electrically small size}. The analysis shows that the latter achieves a higher SE only if a larger number of RF chains is used.}

\subsection{Paper outline and notation}
This paper is organized as follows. Section~\ref{sec:system_model} revisits the system model and capacity achieving scheme of MIMO communications, with particular emphasis on LoS propagations. In Section~\ref{sec:system_model_EM}, first principles of wave propagation are used to introduce an electromagnetic MIMO representation of the communication
problem between two volumes of arbitrary shape and position. In Section~\ref{sec:WDM}, the developed representation is applied to two parallel one-dimensional electromagnetic apertures and used to propose a WDM scheme, which can be seen as the spatial counterpart of OFDM. 
The spectral efficiency of WDM is evaluated in Section~\ref{sec:spectral_efficiency} with different transceiver architectures. Some conclusions and future research directions are discussed in Section~\ref{sec:conclusions}.

We use $\rank({\bf A})$ to denote the rank of matrix $\bf {A}$. For a given vector ${\bf p}$, $\hat {\bf p}$ is a unit vector along its direction and $||{\bf p}||$ denotes its magnitude. $\nabla \times$ denotes the curl operation, $\delta(x)$ the Dirac delta function, and $\floor*{x}$ the greatest integer less than or equal to $x$.

\section{MIMO Communications}  \label{sec:system_model}
A narrowband time-invariant communication system equipped with $N_s$ antennas at the source and $N_r$ antennas at the receiver can be described, at any arbitrary symbol time, by the following discrete-time input-output relation~\cite{TseBook}:
\begin{equation}\label{eq:MIMO_channel}
{\bf y} = {\bf H}{\bf x} + {\bf z}
\end{equation} 
where ${\bf y}\in \mathbb{C}^{N_r}$ and ${\bf x}\in \mathbb{C}^{N_s}$ denote the received and transmitted signal vectors, respectively. The vector ${\bf x}$ must satisfy $\mathbb{E}\{{{\bf x}^{\Htran}{\bf x}}\}\le P$ to constrain the total transmit power. Also, ${\bf z}\sim \mathcal{N}_\mathbb{C}({\bf 0}, {\bf C})$ is {additive Gaussian noise of electromagnetic and hardware nature, with zero-mean and covariance matrix $\bf C$,} and ${\bf H}\in \mathbb{C}^{N_r \times N_s}$ is the MIMO channel matrix. Here, the entry $H_{nm}$ represents the gain between the $m$th source antenna and the $n$th receive antenna. 

\subsection{Capacity and number of degrees-of-freedom}\label{sec:MIMO-capacity}
The capacity of the MIMO channel~\eqref{eq:MIMO_channel} is computed as follows~\cite{TseBook}.
The vector $\bf y$ is premultiplied by ${\bf L}^{-1}$, with ${\bf C}= {\bf L} {\bf L}^{\Htran}$, to obtain ${\bf L}^{-1}{\bf y}$. Let $\widetilde{\bf H}= \widetilde{\bf U}\widetilde{\boldsymbol \Lambda}\widetilde{\bf V}^{\Htran}$ be the SVD of $\widetilde{\bf H}={\bf L}^{-1}{\bf H} $.
Hence, $\widetilde {\bf y} = \widetilde{\bf U}^{\Htran}{\bf L}^{-1}{\bf y}$ reduces to $\widetilde {\bf y} = \widetilde{\boldsymbol \Lambda}\widetilde{\bf x} + \widetilde{{\bf z}}$
where ${\widetilde{\bf x} = \widetilde{\bf V}^{\Htran}{\bf x}}$ and $\widetilde{{\bf z}} = \widetilde{\bf U}^{\Htran}{\bf L}^{-1}{{\bf z}}$ has independent and identically distributed Gaussian entries, i.e., $ \widetilde z_n \sim \mathcal{N}(0,1)$. In scalar form, we have
\begin{align}\label{eq:MIMO_channel_equivalent_scalar}
\widetilde y_n = \widetilde\lambda _n \widetilde{x}_n + \widetilde z_n, \quad n=1,\ldots,\rank(\widetilde{\bf H})
\end{align}
where $\widetilde\lambda_n$ are the non-zero singular values. The MIMO channel is decomposed into $\rank(\widetilde{\bf H})$ non-interfering and independent single-input single-output channels.
Hence, the capacity is~\cite{Telatar}
\begin{align}\label{eq:MIMO-capacity}
C =\sum_{n=1}^{\rank(\widetilde{\bf H})}\log_2 \left(1 + p_n^\star\widetilde \lambda_n^2\right) 
\end{align}
where $p_n^\star$ are the waterfilling solutions:
\begin{align}\label{eq:water-filling}
{p_n^\star = \max\left(0,\mu - \frac{1}{\widetilde\lambda_n^2}\right)}
\end{align}
with $\mu$ being such that the transmit power constraint is satisfied, i.e., $\sum_{n=1}^{\rank(\widetilde{\bf H})} p_n^\star = P$. Each non-zero $\widetilde\lambda_n$ corresponds to a \emph{communication mode} of the channel and can potentially be used to transmit a data stream. The number of communication modes determines the channel DoF~\cite{TseBook}.

\subsection{Line-of-sight channels}
{The simplest channel model for ${\bf H}$ in high-frequency MIMO communications relies on LoS propagation and assume that the source array is located in the \emph{far-field of the receive array}. This condition is known as planar wavefront approximation (e.g.,~\cite{TseBook,Foschini-1999}) and refers to the propagation range at which the direction of arrival of the signal and the channel gain are approximately the same for all the array elements.} In this case, ${\bf H}$ has rank one and a single communication mode with $\lambda_1 = \sqrt{\beta N_rN_s}$, where $\beta$ is the path loss. The capacity is~\cite{TseBook}:
\begin{align}\label{eq:capacity-LoS}
C =\log_2 \left(1 + N_rN_s \frac{\beta P}{\sigma_1^2}\right)
\end{align}
where $\sigma_1^2$ denotes the noise power over the single DoF. Hence, under the far-field LoS assumption, a MIMO system achieves the full array gain $N_rN_s$ but can only transmit a single data stream. This is a direct consequence of the planar wavefront approximation and holds true for any arbitrary array geometry, e.g., linear and planar arrays. {When the distance between the transmitter and receiver is of the same order as the size of arrays, the antenna elements experience different propagation distances and angles. The planar wavefront approximation  breaks down and spherical wave modelling is needed~\cite{Jeng-Shiann-2005,Bohagen2007,Bohagen-2009,Madhow-2011,Heedong-2021}}. In this case, uniform linear arrays can give rise to a channel with $N$ all-equal singular values, provided that $N_r = N_s = N$ and the antenna spacing is $\Delta = \sqrt{\lambda d/N}$~\cite{Madhow-2011}. Similar conditions for other array geometries can also be found, e.g.,~\cite{Bohagen-2005}.

To capture the essence of MIMO communications in different electromagnetic regimes, in the next section we introduce a general MIMO representation of a communication system between two volumes of arbitrary shape and position that is valid for any known channel medium. {Later on, it will be particularized for the setup in Fig.~1.}

\section{Electromagnetic wave communication problem}  \label{sec:system_model_EM}



Consider two volumes of arbitrary shape and position that communicate through an infinite and homogeneous medium, which is characterized by the scalar frequency- and position-independent permeability $\mu_0 = 4\pi \times 10^{-7}$\,[H/m].
An electric current density ${\bf j}(\vect{s},t)$ {at any arbitrary source point ${\bf s}$, inside the source volume $V_s$,} generates an electric field ${\bf e}(\vect{r},t)$ in [V/m] at a generic location  $\vect{r}$ of the receiving volume $V_r$. We consider only monochromatic sources and electric fields, i.e.,
\begin{equation}
{\bf j}(\vect{s},t) = \Re\{{{\bf j}(\vect{s}) e^{-\imagunit \omega t}}\} \quad \text{and} \quad {\bf e}(\vect{r}, t) = \Re\{{\bf e}(\vect{r}) e^{-\imagunit \omega t}\}.
\end{equation}
In this case, Maxwell's equations can be written in terms of the phasor amplitudes ${\bf j}(\vect{s})$ and ${\bf e}(\vect{r})$. 
The sources are assumed to be bounded, so that the current density satisfies 
\begin{equation}\label{eq:power-constraint}
\int_{V_s} \|{\bf j}(\vect{s})\|^2 d {\bf s} < \infty
\end{equation}
which guarantees a bounded radiated power; see Section III-D.

The electric field ${\bf y}({\bf r})$ observed inside volume $V_r$ is the sum of the information-carrying electric field ${\bf e}(\vect{r})$ and a {random noise field} ${\bf n}(\vect{r})$, i.e.,
\begin{align} \label{eq:IO}
{\bf y}(\vect{r})  = {\bf e}(\vect{r}) + {\bf n}(\vect{r})\quad \vect{r}\in V_r
\end{align}
where ${\bf n}(\vect{r})$ accounts for the EMI produced by electromagnetic waves that are not generated by the input source~\cite{Wallace2008}. In Cartesian coordinates, ${\bf e}(\vect{r}) = e_x(\vect{r}) \hat {\bf x} + e_y(\vect{r}) \hat {\bf y} + e_{z}(\vect{r}) \hat {\bf z}$ and ${\bf n}(\vect{r}) = n_x(\vect{r}) \hat {\bf x} + n_y(\vect{r}) \hat {\bf y} + n_x(\vect{r}) \hat {\bf z}$, where $\hat {\bf x}, \hat {\bf y}$ and $\hat {\bf z}$ are the unit vectors, independent of position ${\bf r}$.

\subsection{The Green function and signal-carrying electric field}
In a homogeneous, isotropic medium the electric field ${\bf e}(\vect{r})$ locally obeys the vector wave equation~\cite[Eq.~(1.3.44)]{ChewBook}
\begin{equation} \label{inhomo_Helm_volume}
\nabla \times \nabla \times {\bf e}(\vect{r})-\kappa^2 {\bf e}(\vect{r})=\imagunit \omega \mu_0 {\bf j}(\vect{\vect{r}})=\imagunit \kappa Z_0 {\bf j}(\vect{\vect{r}})
\end{equation}
where $\kappa=\omega/c=2\pi/\lambda$ is the wavenumber (with $c$ being the speed of light) and $Z_{0}$ is the characteristic impedance of the medium (in vacuum, $Z_0 =\mu_0c = 376.73$\,[Ohm]). In an unbounded medium the solution to~\eqref{inhomo_Helm_volume} is given by~\cite[Eq.~(1.3.53)]{ChewBook}
\begin{equation}
\label{FieldCurrent}
{\bf e}(\vect{r})=\imagunit \kappa Z_0 \int_{V_s} {\bf g}(\vect{r},\vect{s}) {\bf j}(\vect{s}) \, d\vect{s}, \quad \vect{r}\in V_r
\end{equation}
where \cite[Eq.~(1.3.51)]{ChewBook}
\begin{equation}
\label{DyadicGF.1}
{\bf g}(\vect{r},\vect{s})= \frac{1}{4 \pi}  \left( \mathbf{I} +\dfrac{\nabla_{\vect r}\nabla^{\Htran}_{\vect r}}{\kappa^2}\right)\frac{e^{\imagunit \kappa \|\vect{r}-\vect{s}\|}}{\|\vect{r}-\vect{s}\|}
\end{equation}
is the Green's function for the electric field. In free-space, under the far-field approximation\footnote{The far-field approximation in electromagnetic propagation corresponds to the radiated field that falls off inversely as the distance apart $\|\vect{r}-\vect{s}\|$. Hence, its power follows the inverse square law.} (i.e. $ \|\vect{r}-\vect{s}\| \gg \lambda$),~\eqref{DyadicGF.1} can be approximated as~\cite{Poon}
\begin{equation}
\label{DyadicGF.2}
\begin{split}
\vect g(\mathbf{r},\mathbf{s}) \approx \frac{1}{4 \pi} \frac{e^{ \imagunit \kappa \|\vect{r}-\vect{s}\|}}{\|\vect{r}-\vect{s}\|}  \left(\mathbf{I}-\widehat{\bf p} \widehat{\bf p}^{\Htran}\right)\end{split}
\end{equation}
where $\widehat{\bf p} = {\bf p}/||{\bf p}||$ and $\bf{p}=\bf{r}-\bf{s}$.
\begin{figure*}[t!]

	\begin{overpic}[width=2\columnwidth,tics=10]{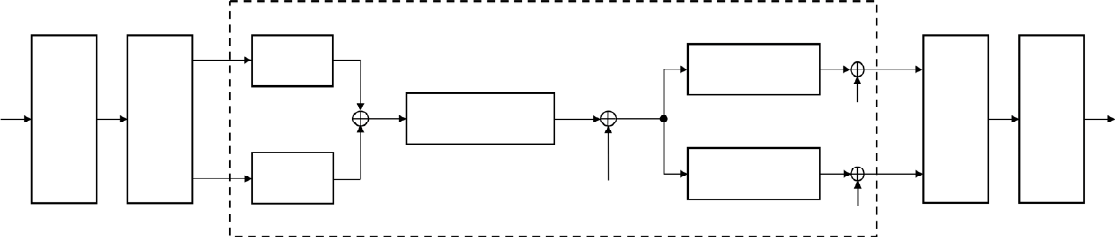}
		\put(-3,12){\small{data}}
		\put(5,6.6){\rotatebox{90}{\small{Waterfilling}}}
	\put(17.7,17){\small{$x_1$}}
	\put(17.7,6.5){\small{$x_N$}}
	\put(74,10){{\footnotesize$z_1^{({\rm hdw})}$}}
	\put(74,1){{\footnotesize$z_N^{({\rm hdw})}$}}
	\put(79.4,16){\small{$y_1$}}
	\put(79.4,6.5){\small{$y_N$}}
	\put(-2,10){\small{$\tilde{\bf x}$}}
	\put(13,10){{$\widetilde {\bf V}$}}
	\put(24,15.7){\small{$\boldsymbol{\phi}_1({\bf s})$}}
	\put(24,5){\small{$\boldsymbol{\phi}_N({\bf s})$}}
	\put(37.2,10.2){\footnotesize{$ \int_{V_s} \!{\bf g}(\vect{r},\vect{s}) {\bf x}(\vect{s})  d\vect{s}$}}
	\put(61.7,15.){\footnotesize{$\int_{\small{V_r}} \!\!\!\boldsymbol{\psi}_1^{\Htran}({\bf r}){\bf y}({\bf r})d{\bf r}$}}
	\put(61.7,5.5){\footnotesize{$\int_{\!\small{V_r}} \!\!\!\boldsymbol{\psi}_N^{\Htran}({\bf r}){\bf y}({\bf r})d{\bf r}$}}
	\put(53,3){\small{${\bf n}({\bf r})$}}
	\put(50.5,12.5){\small{${\bf e}({\bf r})$}}
	\put(55.5,12.5){\small{${\bf y}({\bf r})$}}
	\put(33,12.5){\small{${\bf x}({\bf s})$}}
	\put(99,14){\small{received}}
	\put(100,12){\small{data}}
		\put(101,10){\small{$\tilde{\bf y}$}}
		\put(84,10){{${\bf L}^{-1}$}}
		\put(93,10){{$\widetilde {\bf U}^{\Htran}$}}
			\put(5,0){\small{Pre-processing}}
	\put(84,0){\small{Post-processing}}
	\put(35,22){\small{{Electromagnetic MIMO channel}}}
\end{overpic} \vspace{0.3cm}
	\caption{Optimal transceiver architecture of the electromagnetic MIMO channel for any choice of vectors $\{{\boldsymbol\phi}_m(\vect{s}); m=1,\ldots,N\}$ and $\{{\boldsymbol \psi}_n(\vect{r}); n=1,\ldots,N\}$. The pre-processing and post-processing operations are performed in the digital domain.}
	\label{figure_optimal_transceiver} 
\end{figure*}
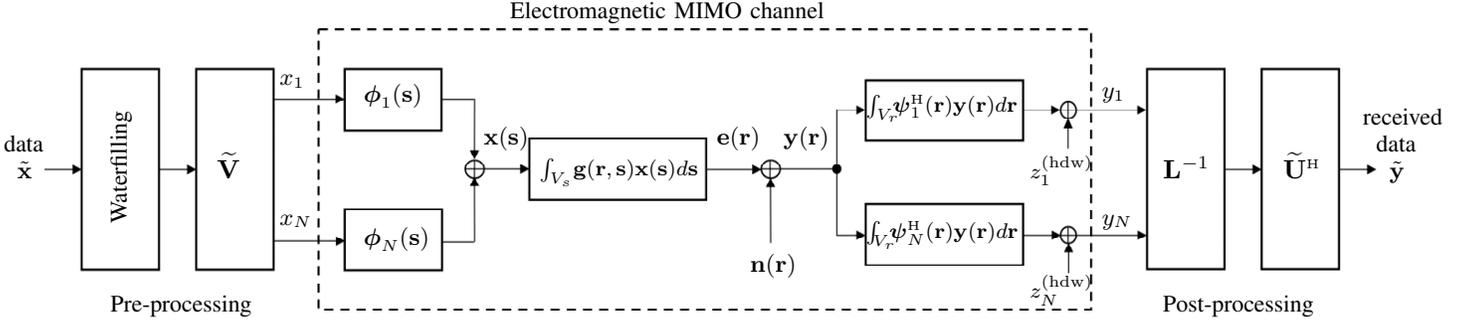

\subsection{The EMI field}\label{sec:noise_field}
{The EMI field} ${\bf n}(\vect{r})$ is produced by the incoming electromagnetic waves that are not generated by the source. Suppose that they are generated in the far-field of the receiving volume. Hence, each electromagnetic wave can be modelled as a plane wave that reaches the receiving volume from an arbitrary elevation angle $\theta_r \in[0,\pi)$ and an arbitrary azimuth angle $\varphi_r\in[-\pi,\pi)$. The EMI field ${\bf n}(\vect{r})$ is thus (e.g.,~\cite{Wallace2008})
\begin{align}\label{eq:noise_r}
{\bf n}(\vect{r}) = \int_{-\pi}^{\pi} \int_{0}^{\pi} {\bf n}(\vect{r}, \theta_r, \varphi_r)  d \theta_r d \varphi_r 
\end{align}
where ${\bf n}(\vect{r}, \theta_r, \varphi_r)$ is defined as
\begin{align}\label{eq:noise_angle}
{\bf n}(\vect{r}, \theta_r, \varphi_r) = {\bf a}(\theta_r, \varphi_r)e^{\imagunit {\boldsymbol \kappa}^{\Ttran}(\theta_r, \varphi_r)\vect{r}}.
\end{align}
Here, $\boldsymbol{\kappa}  (\theta_r,\varphi_r) =\frac{2\pi}{\lambda} \left[\cos \varphi_r \sin \theta_r, \sin \varphi_r \sin \theta_r,\cos \theta_r\right]^{\Ttran}$
is the wave vector that describes the phase variation of the plane wave with respect to the three Cartesian coordinates at the receiving volume, ${\bf a}(\theta_r, \varphi_r)$ is a zero-mean, complex-Gaussian random process with 
\begin{align}\label{eq:noise_coefficients}
\mathbb{E}\{{\bf a}(\theta_r, \varphi_r){\bf a}^{\Htran}(\theta_r^\prime, \varphi_r^\prime)\} = {\sigma}^2_{{{\rm emi}}}f(\theta_r, \varphi_r) {\bf I}_3\delta(\varphi_r-\varphi_r^\prime)\delta(\theta_r-\theta_r^\prime)
\end{align}
and ${\sigma}^2_{{{\rm emi}}}f(\theta_r, \varphi_r)$ denotes the {EMI} power angular density, with ${\sigma}^2_{{{\rm emi}}}$ being measured in V$^2$/m$^2$. From~\eqref{eq:noise_r},  using~\eqref{eq:noise_angle} and \eqref{eq:noise_coefficients}, we have 
\begin{align}
\mathbb{E}\{{\bf n}(\vect{r}^\prime){\bf n}^{\Htran}(\vect{r} + \vect{r}^\prime)\} = {\sigma}^2_{{{\rm emi}}} \rho(\vect{r}){\bf I}_3\label{eq:correlation_1}
\end{align}
with
\begin{align}\label{eq:correlation}
\rho(\vect{r}) = \int_{-\pi}^{\pi} \int_{0}^{\pi}   f(\theta_r, \varphi_r) e^{-\imagunit {\boldsymbol \kappa}^{\Ttran}(\theta_r, \varphi_r)\vect{r} } d \theta_r d \varphi_r.
\end{align}
The above model is valid for any $f(\theta_r, \varphi_r)$. In the case of electromagnetic waves not generated
by the source, we can reasonably assume that they are uniformly distributed in the entire angular domain, given by $\theta_r\in [0,\pi )$ and $\varphi_r\in [ -\pi,\pi)$. This corresponds to an isotropic propagation condition, characterized by
\begin{align}\label{eq:pdf_angular}
f(\theta_r,\varphi_r) = \frac{\sin \theta_r}{4 \pi} \quad \theta_r\in [0,\pi ),\varphi_r\in [ -\pi,\pi).
\end{align}
Plugging~\eqref{eq:pdf_angular} into~\eqref{eq:correlation} yields 
\begin{align}
\rho(\vect{r})=  \int_{-\pi}^{\pi} \int_{0}^{\pi}  \frac{\sin \theta_r}{4 \pi}e^{-\imagunit {\boldsymbol \kappa}^{\Ttran}(\theta_r, \varphi_r)\vect{r} }
d \theta_r d \varphi_r  =\sinc \left(\frac{2||{\bf r}||}{\lambda}\right).\label{rho_r}
\end{align}

\begin{remark}In MIMO communications, the noise samples in~\eqref{eq:MIMO_channel} {of hardware nature} are typically modelled as independent zero-mean and circularly-symmetric Gaussian random variables. This is not generally the case for noise samples of electromagnetic nature as it follows from~\eqref{eq:correlation}. Under the isotropic condition,~\eqref{rho_r} shows that it happens only when $||{\bf r}|| = i \lambda/2$,  with $i \in \mathbb{Z}$, i.e., samples of ${\bf n}(\vect{r})$ are taken along a straight line at a spacing of an integer multiple of $\lambda/2$.\end{remark}

\subsection{The electromagnetic MIMO channel}
Assume that the current density ${\bf j}({\bf s})$ is expanded using a set of orthonormal vectors $\{{\boldsymbol\phi}_m(\vect{s}); m=1,\ldots,N\}$ with $N$ being the dimension of the input signal space. Accordingly, we can write
\begin{align}\label{eq:input_series}
{\bf j}(\vect{s}) = \sum_{m=1}^N \xi_m {\boldsymbol\phi}_m(\vect{s})
\end{align}
where the coefficient $\xi_m$ satisfies
\begin{align}
\xi_m = \int_{V_s} {\boldsymbol\phi}_m^{\Htran}(\vect{s}) {\bf j}(\vect{s}) d\vect{s}.
\end{align}
The received field ${\bf y}(\vect{r})$ in~\eqref{eq:IO} is projected onto an output space of dimension $N$, spanned by a set of receive orthogonal basis functions $\{{\boldsymbol \psi}_n(\vect{r}); n=1,\ldots,N\}$. 
Taking into account the noise introduced by the RF chains at the receiver, the spatial samples $\{y_n; n =1,\ldots,N\}$ are given by
\begin{align}\label{eq:y_n}
y_n = \int_{V_r}  {\boldsymbol \psi}_n^{\Htran}(\vect{r}) {\bf y}(\vect{r}) d\vect{r} + {z_n^{({\rm hdw})}}
\end{align}
{where $z_n^{({\rm hdw})}$ is the noise of hardware nature. We model $\{z_n^{({\rm hdw})}; n =1,\ldots,N\}$ as independent circularly-symmetric Gaussian random variables with variance $\sigma^2_{{\rm hdw}}$.} 
Using \eqref{eq:input_series}--\eqref{eq:y_n}, the input-output relationship takes the form
\begin{align}\label{eq:IO-discrete}
y_n &= \sum_{m=1}^N H_{nm} x_m + z_n \quad \quad n=1,\ldots,N
\end{align}
where 
\begin{align}\label{eq:signal_samples}
x_m = \imagunit \kappa Z_0 \xi_m
\end{align}
are the effective input samples, ${z_n =  z_n^{({\rm emi})} + z_n^{({\rm hdw})}}$ with
\begin{align}\label{eq:noise_samples}
z_n^{({\rm emi})} = \int_{V_r} {\boldsymbol \psi}_n^{\Htran}(\vect{r}) {\bf n}(\vect{r})  d\vect{r}
\end{align}
while
\begin{align}
H_{nm} =\int_{V_r}\int_{V_s} {\boldsymbol \psi}_n^{\Htran}(\vect{r}) {\bf g}({\bf r}, {\bf s}) {\boldsymbol \phi}_m(\vect{s}) d \vect{r} d \vect{s}.\label{eq:H_ni}
\end{align}
Letting ${\bf y} = [y_1,\ldots,y_N]^{\Ttran}$ and ${\bf x} = [x_1,\ldots,x_N]^{\Ttran}$, we may rewrite~\eqref{eq:IO-discrete} in the same matrix form of~\eqref{eq:MIMO_channel}
where ${\bf H}\in \mathbb{C}^{N\times N}$ is the channel matrix and 
${{\bf z}}=[z_1,\ldots,z_N]^{\Ttran}\sim \mathcal{N}_{\mathbb{C}} ({\bf 0}_N, {\bf C})$, with {${\bf C} = \sigma^2_{{\rm emi}}{\bf R} + {\sigma^2_{{\rm hdw}}}{\bf I}_N$} and
\begin{align}\label{eq:correlation_matrix}
\left[ {\bf R}\right]_{nm}= \iint_{V_r}  \rho(\vect{r}-\vect{r}^\prime) {\boldsymbol \psi}_n^{\Htran}(\vect{r}){\boldsymbol \psi}_m(\vect{r}^\prime) d\vect{r}d\vect{r}^\prime
\end{align}
as it follows from~\eqref{eq:noise_samples}, by using~\eqref{eq:correlation_1} and \eqref{rho_r}. Unlike~\eqref{eq:MIMO_channel}, where the entries of ${\bf H}$ represent the channel gains from transmit to receive antennas, here $H_{nm}$ represents the \textit{coupling coefficient} between the source mode $m$ and the reception mode $n$~\cite{Miller:00}. The capacity of \eqref{eq:IO-discrete} is thus given by~\eqref{eq:MIMO-capacity} where the communication modes are obtained by computing the singular value decomposition of $\widetilde{\bf H} = {\bf L}^{-1}{\bf H}$, with the entries of ${\bf H}$ given by~\eqref{eq:H_ni} and ${\bf L} {\bf L}^{\Htran}={\bf C}$. The capacity achieving transceiver architecture for \emph{any choice} of vector basis functions $\{{\boldsymbol\phi}_m(\vect{s}); m=1,\ldots,N\}$ and $\{{\boldsymbol \psi}_n(\vect{r}); n=1,\ldots,N\}$ is reported in Fig.~\ref{figure_optimal_transceiver}, where ${\bf x}{({\bf s})} = \imagunit \kappa Z_0 {\bf j}({\bf s})$. Notice that the transceiver operates in two stages. Particularly, the inner stage operates directly at the electromagnetic level (i.e., analog domain) since it is responsible for generating the current density distribution ${\bf j}({\bf s})$ through the use of functions $\{{\boldsymbol\phi}_m(\vect{s}); m=1,\ldots,N\}$ at the source and for weighting the electric field ${\bf y}(\vect{r})$ according to the functions $\{{\boldsymbol \psi}_n(\vect{r}); n=1,\ldots,N\}$ at the receiver. {The outer stage operates in the digital domain and is responsible to generate $\{x_n; n=1,\ldots,N\}$ and to process $\{y_n; n=1,\ldots,N\}$.}

\subsection{The electromagnetic radiated power}\label{sec:electromagnetic_radiated_power}
In signal processing, the quantity on the left-hand-side of \eqref{eq:power-constraint} represents the \textit{energy} of the spatially-continuous signal ${\bf j}({\bf {s}})$. This is different from the electromagnetic radiated power $\mathsf{P_{rad}}$, which can be computed using the Poynting theorem and integrating the radial component of the Poynting vector over a sphere of radius $r \to \infty$  \cite[Ch. 14]{OrfanidisBook}. In Appendix A, we show that $\mathsf{P_{rad}} $ can be upper-bounded as follows
\begin{equation}
\label{UB_Prad}
\mathsf{P_{rad}} \le Q \mathcal{E}_s
\end{equation}
where $\mathcal{E}_s=\int_{V_s} \|{\bf j}(\vect{s})\|^2 d {\bf s}$
and
\begin{equation}
\label{ }
Q=\dfrac{\kappa Z_0}{4 \lambda} \sqrt{\iint_{V_s}|\rho(\vect{s}_1-\vect{s}_2)|^2d\vect{s}_1 d\vect{s}_2}
\end{equation}
with $\rho(\cdot)$ being defined in \eqref{rho_r}. The upper-bound in \eqref{UB_Prad} shows that imposing the constraint \eqref{eq:power-constraint} on the $L^2$ norm of the source ${\bf j}({\bf s})$ amounts to limit the radiated power. On the contrary, bounding $\mathsf{P_{rad}}$ does not necessarily lead to a bounded source $L^2$ norm in \eqref{eq:power-constraint}; e.g., \cite{Marengo2008}.

\section{Wavenumber-Division Multiplexing}\label{sec:WDM}
The optimal approach for the design of the communication system depicted in Fig.~\ref{figure_optimal_transceiver} consists in selecting the elements of the input basis $\{{\boldsymbol\phi}_n(\vect{s}); n=1,\ldots,N\}$ as the eigenfunctions of the channel operator (e.g., \cite{Miller:00}--~\cite{Wallace2008})
\begin{align}\label{eq:K_s}
{\bf K}_s(\vect{s}, {\bf s}^\prime) =   \int_{V_r} {\bf g}^{\Htran}({\bf r}, {\bf s}){\bf g}({\bf r}, {\bf s}^\prime)d \vect{r}.
\end{align}
This means that
\begin{equation}
\label{EigKs}
\gamma_n \boldsymbol{\phi}_n(\vect s) =  \int_{V_s} {\bf K}_s(\vect{s}, {\bf s}^\prime) \boldsymbol{\phi}_n(\vect s')d \vect{s}'
\end{equation}
where $\gamma_n$ is the eigenvalue associated to the eigenfunction $\boldsymbol{\phi}_n(\vect s)$. The elements of the output basis $\{{\boldsymbol \psi}_n(\vect{r}); n=1,\ldots,N\}$ are simply the channel responses to the functions of the input basis, i.e.,
\begin{equation}
\label{psi_ortho}
 {\boldsymbol \psi}_n(\vect{r}) = \int_{V_s} {\bf g}({\bf r}, {\bf s}) {\boldsymbol \phi}_n(\vect{s})d \vect{s}.
\end{equation}
The above approach is \emph{optimal} under the assumption that the noise can be modelled as a spatially-uncorrelated Gaussian random field~\cite{Miller:00,Marengo2008,Wallace2008}. This may not be the case as described in Section~\ref{sec:noise_field}. In addition, we stress that the numerical solution of the eigenfunction problem~\eqref{EigKs} is in general computationally demanding. Moreover, its implementation is \emph{prohibitive} not only at the electromagnetic (i.e., analog) level but also in the digital domain. We can reasonably ask ourselves whether there is a pair of basis for which the communication system becomes simple to implement and with a clear physical interpretation, though \emph{suboptimal}. 
To answer this question, we consider the LoS scenario in Fig.~\ref{figure_linear_arrays} and take inspiration from a communication system operating over a time-domain dispersive (i.e., frequency selective) channel. The communication over this channel is mathematically equivalent to a time-domain MIMO system. By using the OFDM technology, it can  be converted into a frequency-domain communication system that transmits over multiple non-interfering frequency-flat channels. 
The objective of this section is to show that a communication scheme sharing these advantages can be obtained from the electromagnetic MIMO channel. We refer to it as \emph{WDM (wavenumber-division multiplexing)}. 

\begin{remark}
{We stress that the use of Fourier processing in the spatial domain is not novel as it represents the core of  different technologies and methodologies in the communication and electromagnetic literatures, e.g.,~\cite{Sayeed2013a,Zhang2016a}. Our contribution is not represented by the use of Fourier processing but by the end-to-end description and analysis (using standard tools in signal processing and communication theory) of the arising communication scheme.}
\end{remark}

\subsection{Review of orthogonal-frequency-division multiplexing}
Consider a system that employs $N$ subcarriers with frequency spacing $ 1/T$ and indices in the set $\{-(N-1)/2, . . . , (N-1)/2\}$ with $N$ being an odd integer. The base-band equivalent time-domain signal transmitted over the interval $ 0\le t \le T_s$ with $T_s \ge T$ is
\begin{align}\label{eq:0}
x(t) =  \begin{cases}
\frac{1}{\sqrt{T_s}}
\sum\limits_{m=1}^{N}x_me^{\mathrm{j}\frac{2\pi} {T}\big(m-1-\frac{N-1}{2}\big)t}, &   0\le t \le T_s \\
\;\; 0, & \text{elsewhere}
\end{cases}
\end{align}
where $x_m$ is the data associated to the frequency $f_m = \frac{m-1-(N-1)/2}{T}$.
The signal $x(t)$ is sent over a wireless channel with impulse response $g(t)$. Wireless channel responses are causal and incur a finite maximum delay. We thus assume that the support of $g(t)$ is the time interval $0 \le t \le T_g$. The received signal $y(t)$, observed in $T_g \le t \le T_g+T_r $, is given by
\begin{align}\label{eq:convolution}
{y(t) = g(t)\ast x(t) + z(t)}
\end{align}
where {$\ast$ denotes the convolution operation} and $z(t)$ is the thermal noise. Accordingly, the Fourier transform of $y(t)$, evaluated at $f_n$, is written as:
\begin{align}\label{eq:A28.1}
y_n &= \int_{T_g}^{T_g+T_r} y(t) e^{-\mathrm{j}\frac{2\pi} {T}(n-1 - \frac{N-1}{2})t}dt.
\end{align}   
If 
\begin{align}\label{eq:cyclic_convolution}
T_s \ge T+ T_g \quad \text{and} \quad T_r = T
\end{align}
then~\eqref{eq:A28.1} takes the form~\cite{TseBook}:
\begin{align}
y_n= H_n x_n + z_n\label{eq:A28}
\end{align}
where $H_n = G_nT_r/\sqrt{T_s}$,  with $G_n$ being the channel response at frequency $f_n$, and $z_n$  the corresponding noise.
The expression in~\eqref{eq:A28} describes a set of $N$
non-interfering (orthogonal) parallel transmissions with different complex-valued attenuation factors. Orthogonality is achieved by setting $T_s \ge T+ T_g$ in~\eqref{eq:cyclic_convolution}, which amounts to append the so-called cyclic prefix to the transmitted signal so that the convolution in~\eqref{eq:convolution} can be viewed as a cyclic convolution. In the frequency-domain, it becomes the product of the corresponding Fourier transforms and~\eqref{eq:A28} follows easily. {We notice that the \emph{orthogonality condition} in~\eqref{eq:A28} can also be achieved\footnote{This result will be proved (in the spatial domain) in Appendix C for the proposed WDM scheme.}  if the received signal is observed in the interval $[0,T_r)$, and we set}
\begin{align}\label{eq:cyclic_convolution_v1.0}
T_s = T  \quad \text{and} \quad T_r \ge T+ T_g.
\end{align}
{In this case, no cyclic prefix is appended at the transmitter but the orthogonality can still be retrieved by observing the received signal over an extended time interval $T_r \ge T+ T_g$ that includes the support of the channel response.} The input-output relationship is analogous to \eqref{eq:A28} but with 
\begin{align}
H_n = \sqrt{T_s}G_n. 
\end{align}
Although this is not the way classical OFDM operates, the conditions in~\eqref{eq:cyclic_convolution_v1.0} are instrumental for the design and understanding of the WDM communication scheme described below,  {which will rely on the use of a transmitter of fixed size and a receiver of much larger dimension.}

\subsection{System and signal model}
To simplify the analysis and better highlight the similarities with OFDM, we consider the one-dimensional setup illustrated in Fig.~\ref{figure_linear_arrays}, and we make the following assumption.

\begin{assumption}\label{assumption-tx}
The source is a segment with the following coordinates:
\begin{align}
\{(s_x,s_y,s_z): s_x = 0, s_y =0, |s_z| \le L_s/2\}.
\end{align}
The receiver is a segment occupying the following region:
\begin{align}
\{(r_x,r_y,r_z): r_x =d, r_y =0, |r_z|\le L_r/2, \}.
\end{align}
Also, $L_r = \ell L_s$ with $\ell \ge 1$ being an integer.
\end{assumption}
Under Assumption~\ref{assumption-tx}, the electric current density is
\begin{align}\label{eq:tx_signal}
{\bf j} ({\bf s}) = i(s_z)\delta(s_x) \delta(s_y) \hat{\bf z}
\end{align}
where 
\begin{align}\label{eq:current_signal}
i(s_z) = \sum_{m=1}^{N} \xi_m \phi_m(s_z)
\end{align}
 is a current (measured in amperes). In analogy with an OFDM system, we assume that the basis functions $\{\phi_m(s_z); m=1,\ldots,N\}$ take the form
\begin{align}\label{eq:B0}
\phi_m(s_z) =  \begin{cases}
\;\;\frac{1}{\sqrt{L_s}}e^{\imagunit\frac{2\pi} {L_s}\big(m -1 - {\frac{N- 1}{2}}\big)s_z}, &   |s_z|\le \frac{L_s}{2} \\
\;\; 0, & \text{elsewhere}
\end{cases}
\end{align}
such that
\begin{align}\label{eq:tx_signal}
\xi_m =  \int_{-\frac{L_s}{2}}^{\frac{L_s}{2}} i(s_z) \phi_m^*(s_z) ds_z.
\end{align}
\begin{figure}
\centering
\begin{subfigure}{.5\textwidth}
  \centering
  \includegraphics[width=1.\columnwidth]{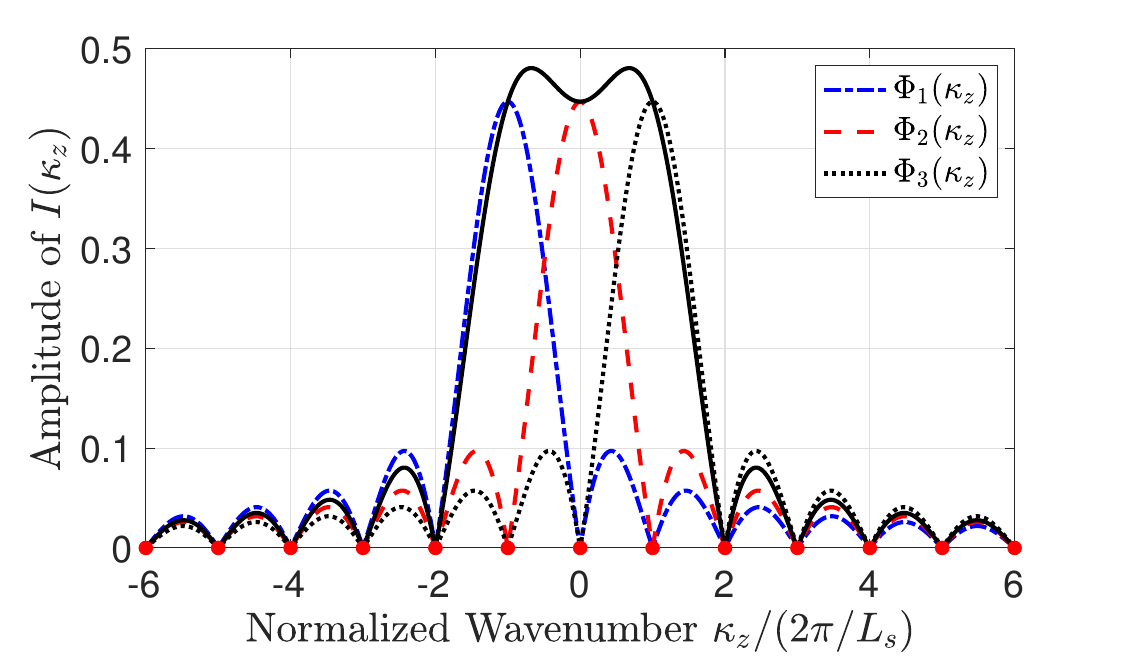}
                \caption{Amplitude of $I(\kappa_z)$.} 
                \label{figure_spatial-DFT_tx} 
\end{subfigure}
\begin{subfigure}{.5\textwidth}
  \centering
  \includegraphics[width=1.\columnwidth]{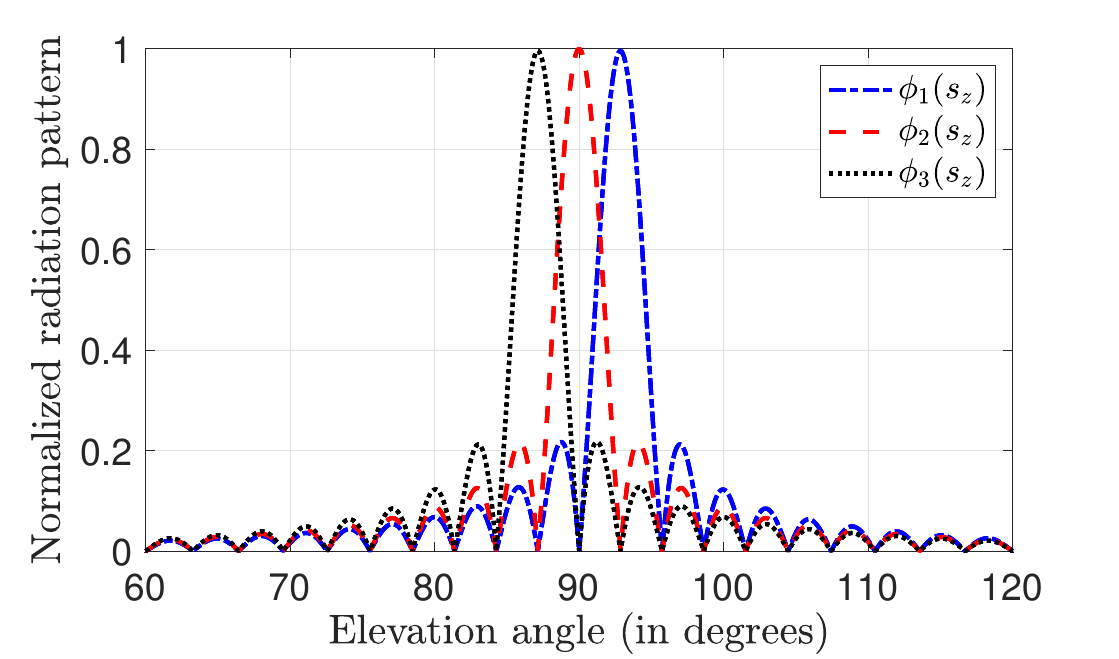}
                \caption{Normalized radiation pattern.} 
                \label{figure_spatial_tx}
\end{subfigure}
        \caption{Behaviour for $L_s =0.2$\,m and $\lambda = 0.01$\,m when $N=3$ and $\xi_{m} =1$ for $m=1,2,3$.}
               \label{fig:green_function}
\end{figure}
The wavenumber Fourier transform of $i(s_z)$ is
\begin{align}\label{eq:X_kappax}
I(\kappa_z) 
= \int_{-\frac{L_s}{2}}^{\frac{L_s}{2}}  \! \!i(s_z) e^{-\imagunit k_zs_z}ds_z = \! \! \!\sum_{m=1}^{N}   \xi_m \Phi_m(\kappa_z )\!\!
\end{align}
where
\begin{align}\notag
\!\!\!\!\Phi_m(\kappa_z ) &=  \int_{-\frac{L_s}{2}}^{\frac{L_s}{2}}  \phi_m(s_z) e^{-\imagunit k_z s_z}ds_z \\&= \sqrt{L_s} {\rm sinc}\left[\left(\frac{\kappa_z}{2\pi}- \frac{m -1 - {(N- 1)/2}}{L_s}\right)L_s\right].\label{eq:Phi_i}
\end{align}
Fig.~\ref{figure_spatial-DFT_tx} plots the amplitude of $I(\kappa_z)$ in~\eqref{eq:X_kappax} for $L_s =0.2$\,m, $\lambda = 0.01$\,m and $N=3$. The distance between the first two nulls of each sinc-function is ${4\pi}/{L_s}$. Fig.~\ref{figure_spatial-DFT_tx} shows that $\Phi_m(\kappa_z)$ has most of its energy in the vicinity of ${ \frac{2\pi}{L_s}(m-1 -{{({N-1})}/{2}})}$. Hence, $I(\kappa_z)$ is practically limited to 
\begin{align}\label{eq:49}
{\left[-\frac{N-1}{2}\frac{2\pi}{L_s}, \frac{N-1}{2}\frac{2\pi}{L_s}\right]}
\end{align}
and thus has an approximate bandwidth of ${ \Omega = (N-1)\frac{\pi}{L_s}}$. {We know that the use of Fourier basis functions produces orthogonal spatial beams, e.g.,~\cite[Ch.~9]{Balanis-2012-antenna,Sayeed2013a,Zhang2016a}. To show this, in Appendix B we derive the electric field generated by $\phi_{m}(s_z)$ along the $z-$axis. Fig.~\ref{figure_spatial_tx} plots the normalized radiation pattern in~\eqref{eq:E.10} as a function of the elevation angle. We see that each $\phi_m(s_z)$ with $m=1,2,3$ radiates towards an angular direction given by
\begin{equation}
\theta_{m} = \cos^{-1}\left(\frac{\lambda}{L_s}\bigg(m -1 - {\frac{N- 1}{2}}\bigg)\right).
\end{equation}
If we define the beamwidth as the angular distance between the first two nulls around $\theta_{m}$, it is equal to $2 \lambda/L_s$ as it follows from~\eqref{eq:E.9} in Appendix B. Hence, in line with classical results, it is inversely proportional to the array length $L_s$.}

At the receiver, ${\bf y}({\bf r})$ is projected onto the space spanned by the vectors ${\boldsymbol{\psi}}_n({\bf r})$ 
given by
\begin{align}\label{eq:B0-rx_1}
{\boldsymbol{\psi}}_n({\bf r}) = \psi_n(r_z) \delta(r_x-d) \delta(r_y){\hat {\bf z}}
\end{align}
where, in analogy with OFDM, 
\begin{align}\label{eq:B0-rx}
\psi_n(r_z) =  \begin{cases}
\;\;e^{\imagunit\frac{2\pi} {L_s}\big(n -1 - {\frac{N- 1}{2}}\big)r_z}, &   |r_z|\le L_r/2 \\
\;\; 0, & \text{elsewhere.}
\end{cases}
\end{align}
Notice that the fundamental spatial frequency of $\psi_n(r_z)$ is $1/L_s$, as at the source side.

In the wavenumber domain, we have that
\begin{align}
\Psi_n(\kappa_z ) 
= {L_r} {\rm sinc}\left[\left(\frac{\kappa_z}{2\pi}-\frac{n -1 - {(N- 1)/2}}{L_s}\right)L_r\right]\label{eq:B1}
\end{align}
where the distance between the first two nulls is now ${4 \pi}/{L_r}$ and thus inversely proportional to $L_r$, rather than $L_s$. From~\eqref{eq:y_n}, by using~\eqref{eq:B0-rx} we obtain
\begin{align}\notag
y_n &= \int_{-\frac{L_r}{2}}^{\frac{L_r}{2}}  \psi_n^*(r_z)  y(d,0,r_z) dr_z + {z_n^{({\rm hdw})}}\\&=\int_{-\frac{L_r}{2}}^{\frac{L_r}{2}} y(d,0,r_z)  e^{-\imagunit\frac{2\pi} {L_s}\big(n -1 -{\frac{{N-1}}{2}}\big)r_z}dr_z + {z_n^{({\rm hdw})}} \label{eq:dft-rx}
\end{align}
where the first term is simply the $n$th coefficient of the wavenumber Fourier series expansion of $y(d,0,r_z) $ taken at $1/L_s$. {Clearly, the projection of ${\bf y}({\bf r})$ onto the space spanned by~\eqref{eq:B0-rx_1} allows us to only process the electric field along the $z-$axis, which is represented by $y(d,0,r_z)$ in~\eqref{eq:dft-rx}. Notice that $y_n$ in~\eqref{eq:dft-rx} is expressed in volts.}

\begin{figure}
\centering
\begin{subfigure}{.5\textwidth}
  \centering
  \includegraphics[width=1.\columnwidth]{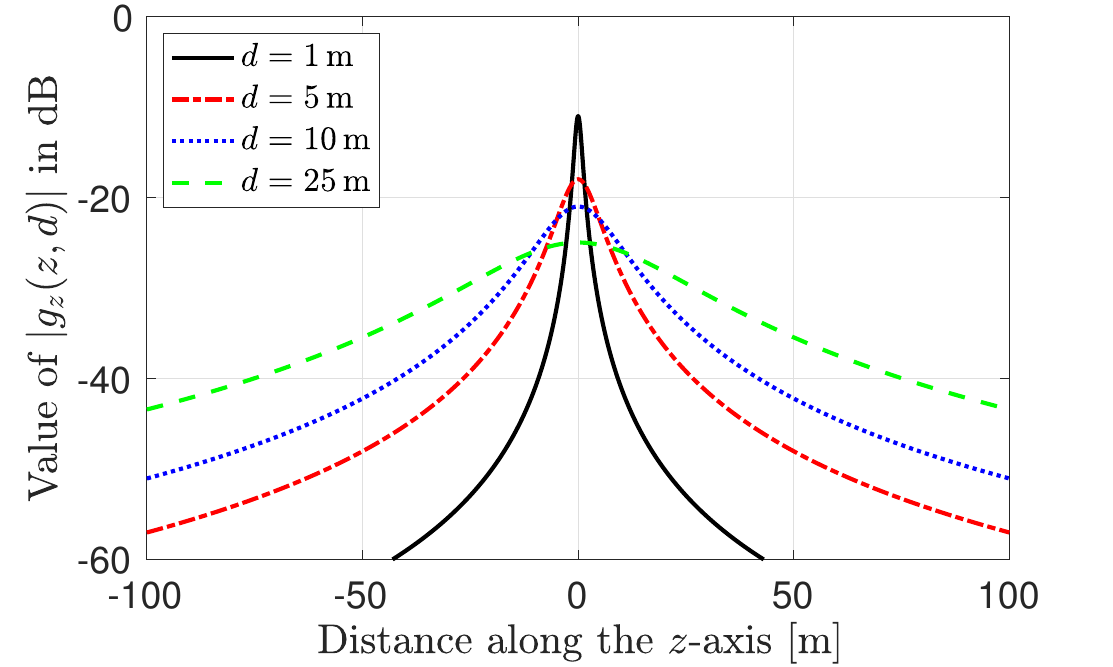}
                \caption{Amplitude of $|g_z(z,d)|$.} 
                \label{fig:spatialgreen_function} 
\end{subfigure}%

\begin{subfigure}{.5\textwidth}
  \centering
  \includegraphics[width=1.\columnwidth]{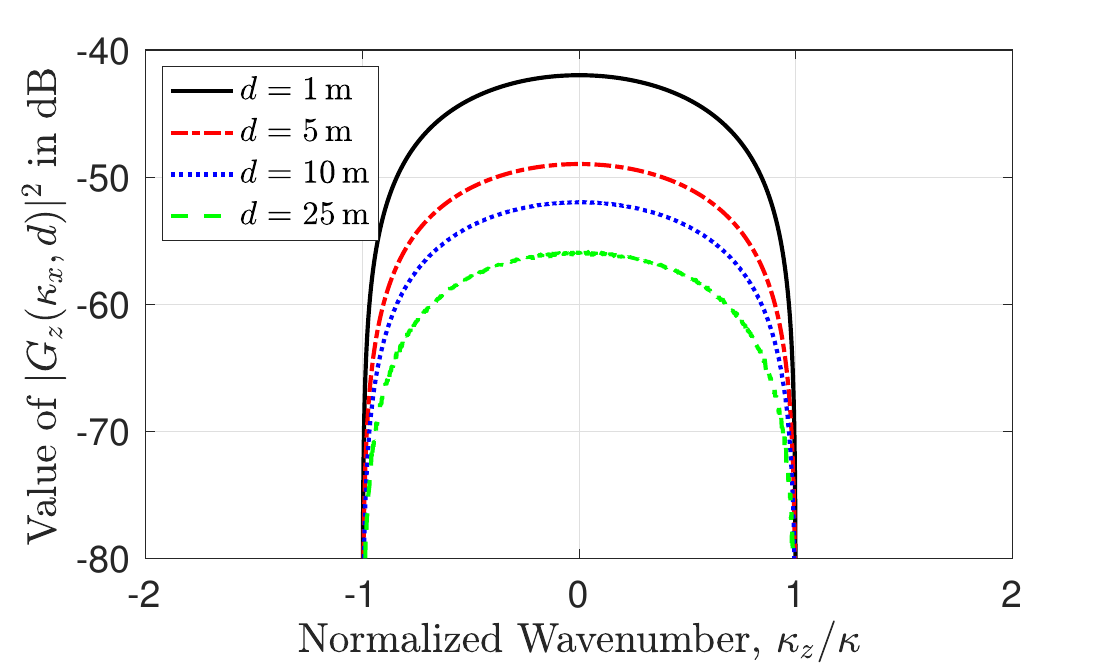}
                \caption{Value of $|G_z(\kappa_z,d)|^2$.} 
                \label{fig:frequencygreen_function}
\end{subfigure}
        \caption{Behaviour of $|g_z(z,d)|$ and $|G_z(\kappa_z,d)|^2$ for $d = 1,5, 10$ and $25$\,m when $\lambda =0.01$\,m (i.e., carrier frequency of $30$ GHz).}
        \label{fig:green_function}
\end{figure}

\subsection{The coupling coefficients}
Under the system model above, in free-space propagation~\eqref{eq:H_ni} reduces to\footnote{{The double integral in~\eqref{eq:H_ni-free-space_linear} can easily be written as a single integral to facilitate its evaluation.}}
\begin{align}\label{eq:H_ni-free-space_linear}
\!\!H_{nm} =  \int_{-\frac{L_r}{2}}^{\frac{L_r}{2}}\int_{-\frac{L_s}{2}}^{\frac{L_s}{2}} \!\!\!\! g_z(r_z-s_z,d)  \psi_n^*(r_z)  \phi_m(s_z) d s_z d r_z
\end{align}
where $g_z(z,d)$ denotes the $(3,3)$th element of ${\bf g}({\bf r}, {\bf s})$ evaluated at $r_x = d$ and $s_x =s_y =r_y = 0$. From~\eqref{DyadicGF.2}, we obtain
\begin{equation}  \label{Green-linear}
g_z(z, d) =    \frac{d^2}{4 \pi} \frac{e^{\imagunit \kappa \sqrt{z^2 + d^2}}}{\left(z^2 + d^2\right)^{3/2}}.
\end{equation}
We call
\begin{align}\label{eq:G_z}
G_z(\kappa_z,d) = \int_{-\infty}^{\infty} g_z(z, d) e^{-\imagunit \kappa_z z} d z
\end{align}
the Fourier transform in the wavenumber domain of $g_z(z, d)$, and hence  we can write
\begin{align}\label{eq:g_xrz}
g_z(z, d) = \frac{1}{2\pi}\int_{-\infty}^{\infty}  G_z(\kappa_z,d) e^{\imagunit \kappa_z z} d \kappa_z.
\end{align}
The shape of $|g_z(z,d)|$ and $|G_z(\kappa_z,d)|^2$ in dB is illustrated in Fig.~\ref{fig:green_function} for $d = 1, 5, 10$ and $25$\,m. Unlike time-domain channels, $g_z(z,d)$ is not causal and has unbounded support. Also, $G_z(\kappa_z,d)$ is band-limited in the interval $|\kappa_z| \le \kappa = 2 \pi/\lambda$. The lower the wavelength $\lambda$, the larger the spatial-frequency bandwidth. The $3-$dB bandwidth of $G_z(\kappa_z,d)$ is approximately $ \alpha \kappa$ with $\alpha = 0.6$.

\begin{figure}
\centering
\begin{subfigure}{.5\textwidth}
  \centering
  \includegraphics[width=1.\columnwidth]{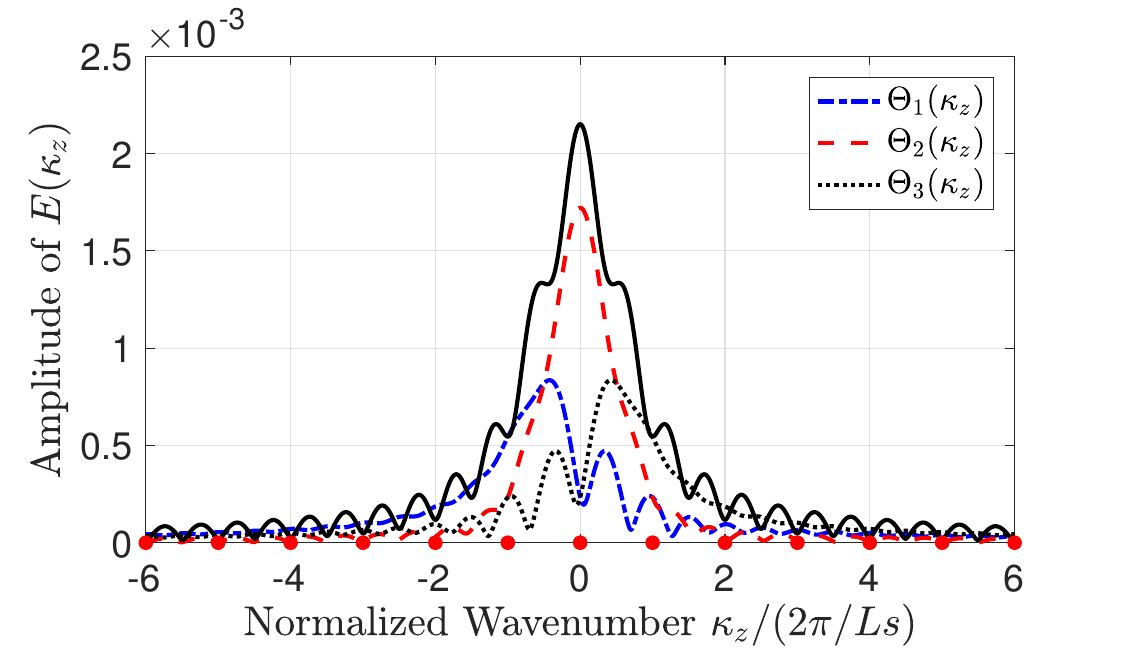}
                 \caption{$L_r= 0.4$\,m.} 
                \label{fig:fig5a} 
\end{subfigure}%

\begin{subfigure}{.5\textwidth}
  \centering
  \includegraphics[width=1.\columnwidth]{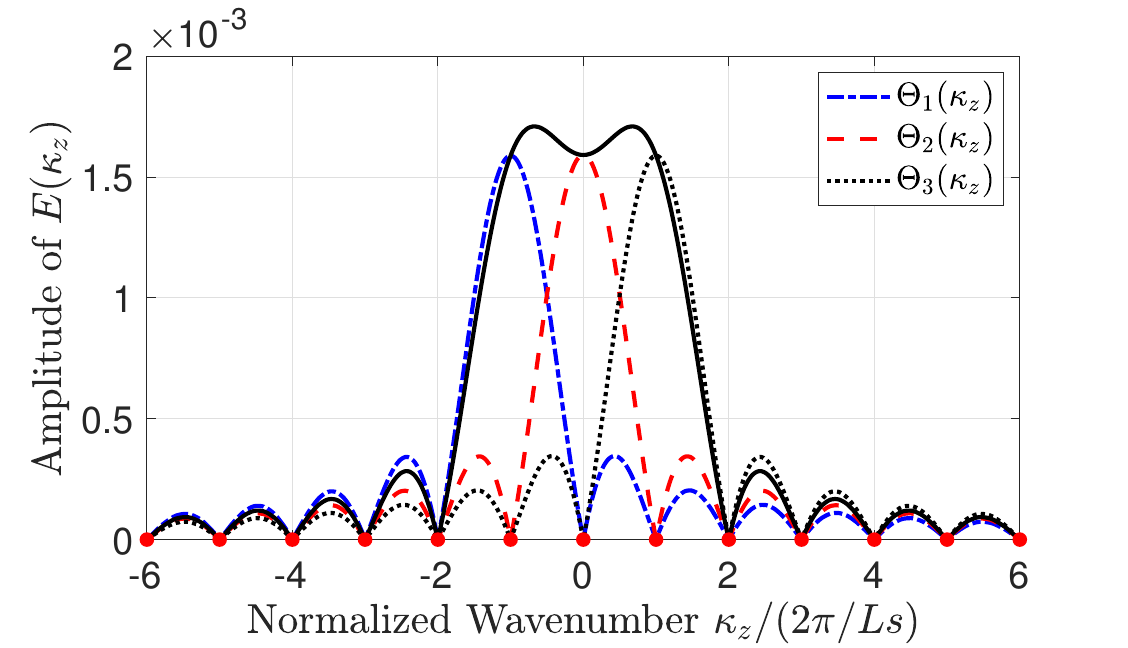}
                \caption{$L_r= 5$\,m.} 
                \label{fig:fig17}
\end{subfigure}
        \caption{Amplitude of $E(\kappa_z)$ and $\Theta_m(\kappa_z)$ for $m=1,2,3$ in the same setup of Fig.~\ref{figure_spatial-DFT_tx} when $d=5$\,m and $L_r=0.4$ and $5$\,m.}
        \label{fig:rx_electric_field}
\end{figure}

To quantify the impact of the propagation channel, we reformulate~\eqref{eq:H_ni-free-space_linear} as follows
\begin{align}\label{eq:H_ni-free-space_linear_eigenfunctions}
\!\!H_{nm} =  \int_{-\frac{L_r}{2}}^{\frac{L_r}{2}}\psi_n^*(r_z) \theta_m(r_z) d r_z
\end{align}
with 
\begin{align}\label{eq:optimal_RX_basis_functions}
 \theta_m(r_z) = \int_{-\frac{L_s}{2}}^{\frac{L_s}{2}} \!\!\!\! g_z(r_z-s_z,d) \phi_m(s_z) d s_z
 \end{align}
being the $m$th element of the basis spanning the electric field at the receiver, i.e., $ e(d,0,r_z)=\sum_{m=1}^N x_m  \theta_m(r_z)$.
 We call $ E(\kappa_z) $ and $ \Theta_m(\kappa_z) $ the Fourier transforms in the wavenumber domain of $ e(d,0,r_z)$ and  $\theta_m(r_z)$, respectively. Fig.~\ref{fig:rx_electric_field} plots their behaviors in the same setup of Fig.~\ref{figure_spatial-DFT_tx} when $d=5$\,m and $L_r=0.4$ and $5$\,m. From Fig.~\ref{fig:fig5a}, we see that when $L_r=0.4$ the spectrum of the electric field is highly distorted compared to that in Fig.~\ref{figure_spatial-DFT_tx}. Moreover, the orthogonality condition among the three communication modes is destroyed. The situation is much different when $L_r=5$\,m. In this case, the electric field is received almost undistorted over the considered wavenumber interval and the three communication modes are barely affected by interference.

\begin{figure}
\centering
\begin{subfigure}{.5\textwidth}
  \centering
  \includegraphics[width=1.\columnwidth]{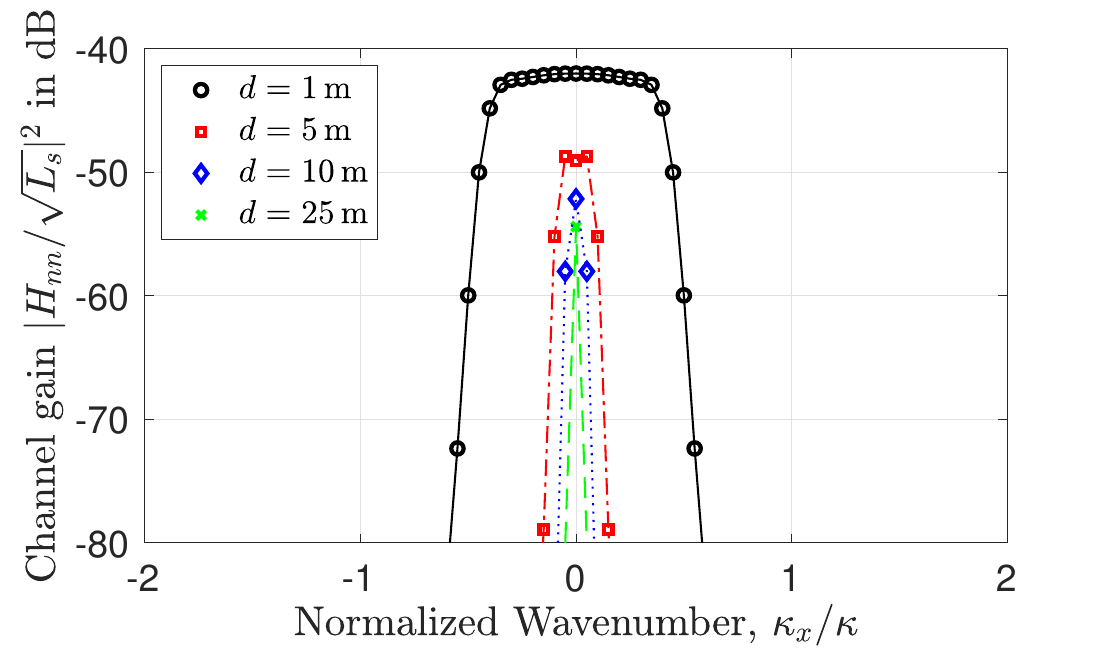}
                \caption{$L_r=1$\,m}
                \label{fig:H_nn_function_1}
\end{subfigure}%

\begin{subfigure}{.5\textwidth}
  \centering
  \includegraphics[width=1.\columnwidth]{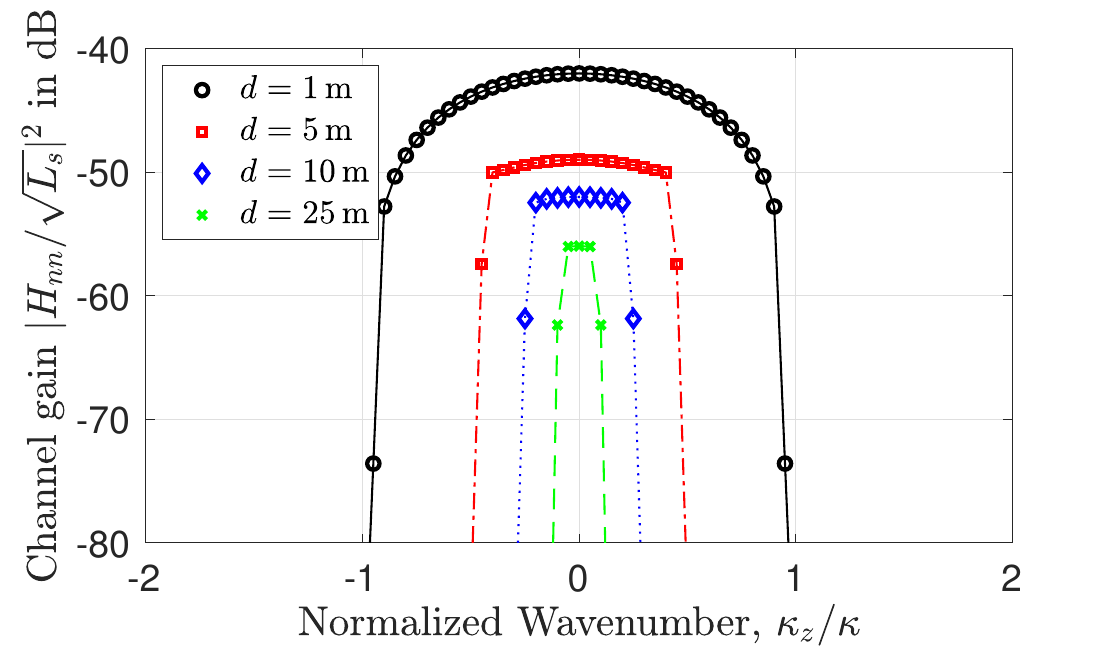}
                \caption{$L_r=5$\,m}
                \label{fig:H_nn_function_2}
\end{subfigure}
        \caption{Behaviour of $|H_{nn}/\sqrt{L_s}|^2$ in~\eqref{eq:coupling-coefficientes-linear} in dB in the same setup of Fig.~\ref{fig:green_function} when $L_r=1$ and $5$\,m.}
        \label{fig:channel_function}
\end{figure}

By using~\eqref{eq:g_xrz}, we may rewrite~\eqref{eq:H_ni-free-space_linear} as
\begin{align}\label{eq:H_ni_frequency_domain_x}
H_{nm} = \frac{ 1}{2\pi}\int G_z(\kappa_z,d)  \Psi_n^*(\kappa_z)  \Phi_m(\kappa_z) d \kappa_z \end{align}
where $\Phi_m(\kappa_z)$ and $\Psi_n^*(\kappa_z)$ are given by~\eqref{eq:Phi_i} and~\eqref{eq:B1}, respectively.  Hence, we obtain~\eqref{eq:coupling-coefficientes-linear} at the top of next page.
\begin{figure*}[t!]
\begin{align}
H_{nm}=  \frac{ {\sqrt{L_s}L_r}}{2\pi} \int_{-\infty}^\infty \!\!\!\!G_z(\kappa_z,d)  \sinc\left[\left(\frac{\kappa_z}{2\pi}-\frac{n-1 - {{{(N-1)}}/2}}{L_s}\right) L_r\right]{\sinc}\left[\left(\frac{\kappa_z}{2\pi}-\frac{m -1- {{{(N-1)}}/2}}{L_s}\right)L_s\right]  d\kappa_z\label{eq:coupling-coefficientes-linear}
\end{align}
\hrule
\end{figure*}

\begin{figure}[t!]
	\centering 
	\begin{overpic}[width=1.\columnwidth,tics=10]{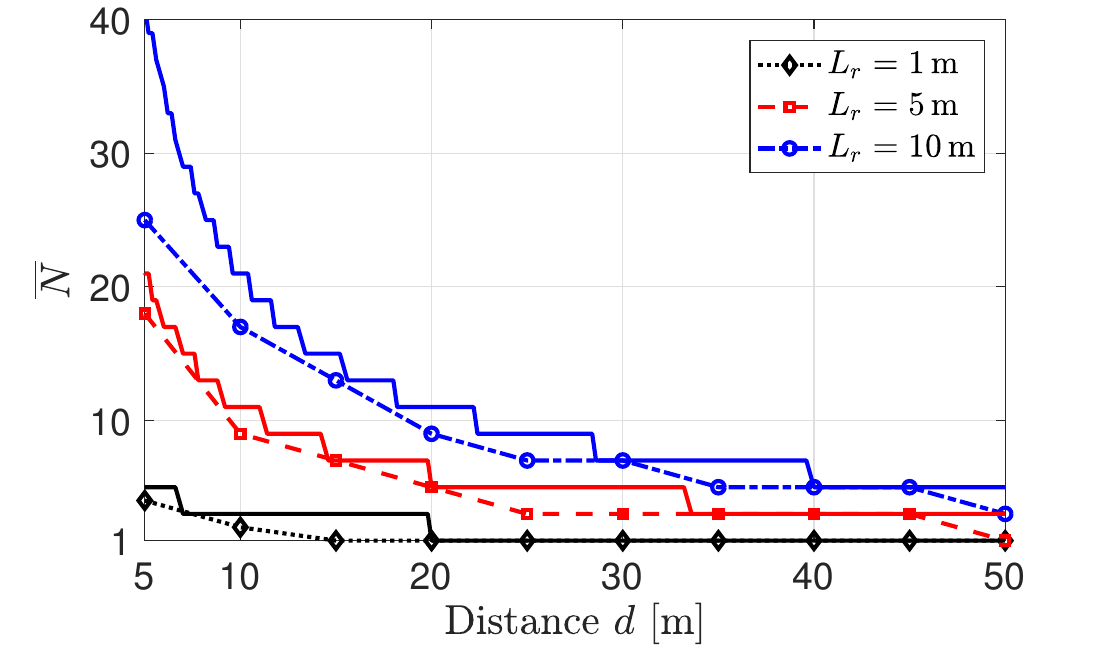}
\end{overpic} 
	\caption{Number $\overline N $ of coupling coefficients $H_{nn}$ that resides in the $3-$dB bandwidth of $G_z\left(\kappa_z,d\right)$. The solid curves are obtained by using~\eqref{eq:paraxial_approximation} with $L_s=0.2$\,m and $\lambda = 0.01$\,m.}
	\label{figure_barN} 
\end{figure}

In Fig.~\ref{fig:channel_function}, we illustrate $|H_{nn}/\sqrt{L_s}|^2$ for the same setup of Fig.~\ref{fig:green_function}, when $L_r=1$ and $10$\,m. As seen, the behaviour of $|H_{nn}/\sqrt{L_s}|^2$ is much different from that of $|G_z\left(\kappa_z,d\right)|^2$ in Fig.~\ref{fig:green_function}. The number of significant coefficients reduces as $d$ increases. This is exemplified in Fig.~\ref{figure_barN} where we plot the number $\overline N $ of coupling coefficients $H_{nn}$ that resides in the $3-$dB bandwidth of $G_z\left(\kappa_z,d\right)$ in Fig.~\ref{fig:frequencygreen_function} for $L_r =1$, $5$ and $10$\,m. We see that if $d=5$\,m then $\overline N = 3$, $17$ and $25$ for $L_r =1$, $5$ and $10$\,m, respectively. These numbers reduce to $\overline N = 1$, $9$ and $16$ when $d=10$\,m. The solid curves are obtained from~\eqref{eq:paraxial_approximation} as $2\floor*{ \frac{1}{2}\frac{L_s L_r}{\lambda d}}+1$ (modified in order to have an odd number of communication modes) 
and represent a good approximation of $\overline N$ only when $d\ge L_r$. This is in agreement with the fact that~\eqref{eq:paraxial_approximation} is valid only in that regime (e.g.,~\cite{Miller:00}).

A comparison between Fig.~\ref{fig:H_nn_function_1} and Fig.~\ref{fig:H_nn_function_2} shows that, for a given distance $d$, the coupling coefficient $|H_{nn}/\sqrt{L_s}|^2$ tends to the sampled version of $|G_z\left(\kappa_z,d\right)|^2$ as $L_r$ increases. In fact, the following result holds true when $L_r$ grows unboundedly.

\begin{corollary}\label{corollary1}
If $L_r\to \infty$ then
\begin{align}\label{eq:orthogonality-condition}
H_{nm}  \to    \left\{{\begin{array}{cc}
    \sqrt{L_s}G_n,& n=m \\
   0& n \ne m \\
  \end{array} } \right.
\end{align}
with $G_n = G_z\left(\frac{2\pi}{L_s}\left(n -1 - {\frac{{N-1}}{2}}\right) , d\right)$.
\end{corollary}

\begin{IEEEproof}
If $L_r\to \infty$ then
\begin{align}
\frac{1}{2 \pi}\Psi_n(\kappa_z ) \to \delta\left(\kappa_z -2\pi \frac{n -1 - {{{(N-1)}}/{2}}}{L_s}\right). 
\end{align}
By using the above result into~\eqref{eq:coupling-coefficientes-linear} yields $ H_{nm}\to  \sqrt{L_s} G_n\sinc\big(n-m\big)$
where $G_n $ is obtained from~\eqref{eq:G_z} by taking $\kappa_z=\frac{2\pi}{L_s}(n -1 -(N -1)/2) $. Since $\sinc\big(n-m\big)$ is $1$ for $n=m$ and $0$ otherwise, the proof is completed.
\end{IEEEproof}
The above corollary shows that asymptotically $H_{nn}/\sqrt{L_s}$ tends to $G_n$. Since $G_z(\kappa_z,d)$ is band-limited with bandwidth $\Omega =2 \pi/\lambda$, from~\eqref{eq:49} the maximum dimension of the input signal space is thus given by
\begin{align}\label{eq:max_number_of_modes}
{N_{\max} = 2\floor*{ \frac{L_s}{\lambda}}+1}
\end{align}
which corresponds to the maximum DoF of the electromagnetic channel in the 1D setup in Assumption 1; e.g.,~\cite{SPAWC}. If the $3-$dB bandwidth is considered, then the approximate dimension reduces to $\floor*{\alpha N_{\max}}$ with $\alpha = 0.6$. 

Recalling that $H_{nm}$ represents the coupling coefficient between source mode $m$ and reception mode $n$, the above corollary shows that the communication modes are decoupled only when $L_r$ grows to infinity, i.e., when an infinitely large receiving segment is used. This is a direct consequence of the fact that $g_z(z,d)$ is not limited in the spatial-domain. However, the following result can be proved.
\begin{corollary}\label{corollary2}
If $|g_z(z,d)|$ is much smaller than $|g_z(0,d)|$ for $|z| \ge z_g$, i.e., 
\begin{align}\label{eq:B31}
|g_z(z,d)| \ll |g_z(0,d)| \quad |z| \ge z_g\end{align}
then
\begin{align}\label{eq:B30}
L_r\ge L_s + z_g
\end{align}
guarantees~\eqref{eq:orthogonality-condition}.
\end{corollary}
\begin{IEEEproof}
See Appendix C.
\end{IEEEproof}

Corollary 2 shows that the orthogonality condition~\eqref{eq:orthogonality-condition} can be achieved if $L_r$ is larger than a predefined value $z_g$ such that $|g_z(z, d)|$ in~\eqref{Green-linear} is negligible for $|z| \ge z_g$. Hence,~\eqref{eq:B30} can be viewed as the spatial counterpart of the time-domain condition~\eqref{eq:cyclic_convolution_v1.0} in OFDM systems. In other words, the extra length $L_r - L_s \ge z_g$ allows to observe the received signal over its entire support in the spatial domain. As proved in Appendix C, this restores the orthogonality among communication modes and basically converts the spatial domain dispersive channel into a set of flat channels in the wavenumber domain. Numerical results will show that the asymptotic orthogonality condition~\eqref{eq:orthogonality-condition} is sufficiently accurate starting from a receiver size $L_r$ of the same order of the source distance $d$. 

\begin{remark}Fig.~\ref{figure_barN} shows that $\overline N$ tends to $1$ as $d$ increases, for any size $L_r$. Notice that when $d$ grows large and $L_r,L_s,\lambda$ are kept fixed, we can approximate $g_z(z, d)$ as $g_z(z, d) \approx   \frac{1}{4 \pi}\frac{e^{\imagunit \kappa {d}}}{d}$,
which corresponds to the Green's function in the far-field of source and receiver. This is valid for the propagation range
at which the direction and channel gain are approximately the
same, i.e., {planar wavefront approximation}. In this case,~\eqref{eq:H_ni-free-space_linear} reduces to
\begin{align}\label{eq:H_ni-free-space_linear_reduced}
\!\!H_{nm} =  \frac{1}{4 \pi} \frac{e^{\imagunit \kappa {d}}}{d} \int_{-\frac{L_r}{2}}^{\frac{L_r}{2}} \psi_n^*(r_x)  d r_x\int_{-\frac{L_s}{2}}^{\frac{L_s}{2}}\phi_m(s_x) d s_x
\end{align}
from which it follows
\begin{align}
 H_{nm} = \left\{{\begin{array}{cc}
    \frac{1}{4 \pi} \frac{e^{\imagunit \kappa {d}}}{d} L_r \sqrt{L_s} ,& n=m=(N-1)/2+1 \\
   0& {\text{otherwise.}} \\
  \end{array} } \right.
\end{align}
In line with classical results, a single DoF is available {under the planar wavefront approximation}.
\end{remark}

\subsection{The {EMI} samples}
From~\eqref{eq:noise_samples}, the {EMI} samples are
\begin{align}\label{eq:noise_samples_1D}
z_n^{({\rm emi})}&= \int_{-\frac{L_r}{2}}^{\frac{L_r}{2}}  \psi_n^*(r_z)  n(d,0,r_z) dr_z 
\end{align}
and the matrix ${\bf R}$ in~\eqref{eq:correlation_matrix} has entries\footnote{{Similar to~\eqref{eq:H_ni-free-space_linear}, the double-integral in~\eqref{eq:correlation_matrices_entries_noise} can easily be written as a single one.}}
\begin{align}\label{eq:correlation_matrices_entries_noise}
\!\!\left[ {\bf R}\right]_{nm} \!\!= \!\iint_{-\frac{L_r}{2}}^{\frac{L_r}{2}} \rho\left({r_z-r_z^\prime}\right)  \psi_n^*(r_z)  \psi_m(r_z^\prime) dr_zdr_z^\prime
\end{align}
{with $\rho\left(z\right) = \int_{-\pi}^{\pi} \int_{0}^{\pi}   f(\theta_r, \varphi_r) e^{-\imagunit { \kappa}_z(\theta_r, \varphi_r)z } d \theta_r d \varphi_r$ as it follows from~\eqref{rho_r} under Assumption 1. As done for~\eqref{eq:H_ni-free-space_linear}, we may rewrite~\eqref{eq:correlation_matrices_entries_noise} as
\begin{align}\label{eq:H_ni_frequency_domain_x_noise}
\left[ {\bf R}\right]_{nm}  &= \frac{ 1}{2\pi}\int_{-\infty}^\infty S_z(\kappa_z)  \Psi_n^*(\kappa_z)  \Psi_m(\kappa_z) d \kappa_z\end{align}
where $S_z(\kappa_z) = \int_{-\infty}^{\infty} \rho(z) e^{-\imagunit \kappa_z z} d z$ is the wavenumber Fourier transform of $\rho(z)$. Clearly, the spatial channel correlation properties of the EMI samples depend on the power angular density $ f(\theta_r, \varphi_r)$. Under the isotropic propagation conditions characterized by~\eqref{eq:pdf_angular}, from~\eqref{rho_r} we have that $\rho(z) = \sinc \left(2\frac{{z}}{\lambda}\right)$ and
\begin{align}\label{eq:S_x}
S_z(\kappa_z) 
= \frac{\pi}{\kappa} {\rm{rect}}\left(\frac{\kappa_z}{2\kappa}\right) = \frac{\lambda}{2} {\rm{rect}}\left(\frac{\kappa_z}{2\kappa}\right)
\end{align}
which is a rectangular function of bandwidth $\kappa$.} Hence, the samples $\{z_n^{(\rm{emi})}\}$ are correlated despite the isotropic nature of the EMI ${\bf n}({\bf r})$. This is a direct consequence of the finite size of $L_r$ that limits the observation interval of ${\bf n}({\bf r})$ in the spatial domain. The following result is obtained. 
\begin{corollary}\label{corollary2}
If $L_r\to \infty$, then
\begin{align}\label{eq:orthogonality-condition-noise}
{\sigma}_{{\rm emi}}^2 \frac{\left[ {\bf R}\right]_{nm}}{L_r}   \to    \left\{{\begin{array}{cc}
     \frac{ {\sigma}_{{\rm emi}}^2}{2 \kappa},& n=m \\
   0& n \ne m \\
  \end{array} } \right.
\end{align}
for $n \le N_{\max} = 2\floor*{ \frac{L_s}{\lambda}}+1$.
\end{corollary}
\begin{IEEEproof}
It follows from~\eqref{eq:H_ni_frequency_domain_x_noise} by using \eqref{eq:S_x} and the same steps in the proof of Corollary~\ref{corollary1}. 
\end{IEEEproof}
The implication of the above corollary is that, as $L_r$ grows large, the covariance matrix ${\bf R}$ becomes diagonal with elements given by $\frac{ L_r}{2 \kappa}$; that is, the samples $\{z_n^{({\rm emi})}\}$ become uncorrelated with a variance that increases unboundedly. In practice, this unbounded behaviour of the noise variance is a consequence of the model in Section~\ref{sec:noise_field}, which becomes not physically meaningful in the asymptotic regime as it results into noise samples of infinite power.
Moreover, it would asymptotically lead to a communication system characterized by a null SE since Corollary~\ref{corollary1} showed that the coupling coefficients tend to a finite quantity as $L_r \to \infty$. However, we anticipate that all this is not an issue since the asymptotic orthogonality condition in Corollary~\ref{corollary1} is sufficiently accurate for $L_r \ge d$. Hence, in scenarios of practical relevance, the variance will be large but always finite, and the {EMI} model from Section~\ref{sec:noise_field} can be safely used.

        \begin{figure}[t!]
	\centering 
	\begin{overpic}[width=1\columnwidth,tics=10]{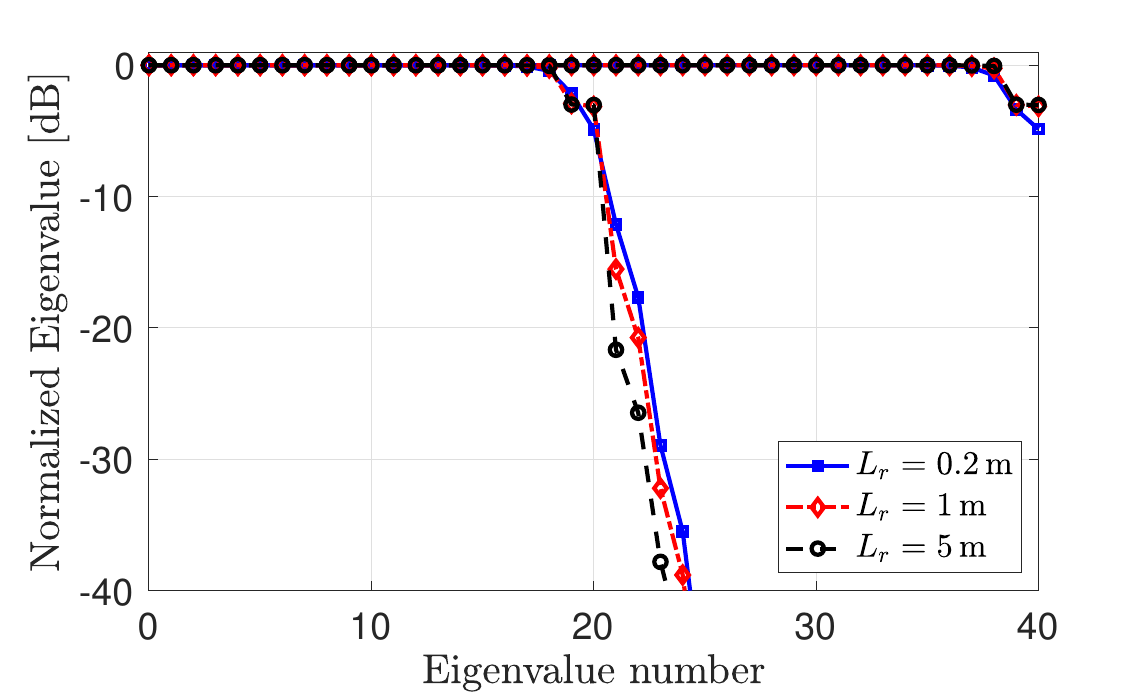}
		\put(60,45){\footnotesize Isotropic EMI}
	\put(68,48){\vector(1,1){5}}
		\put(26,24){\footnotesize Non-isotropic EMI}
	\put(46,28){\vector(1, 1){5}}
\end{overpic} 
	\caption{Normalized eigenvalues in dB of $\bf R$ for $L_s = 0.2$\,m and $\lambda=0.01$\,m. Isotropic and non-isotropic propagation conditions for EMI are considered.}
	        \label{eigenvalues_noise}
\end{figure}
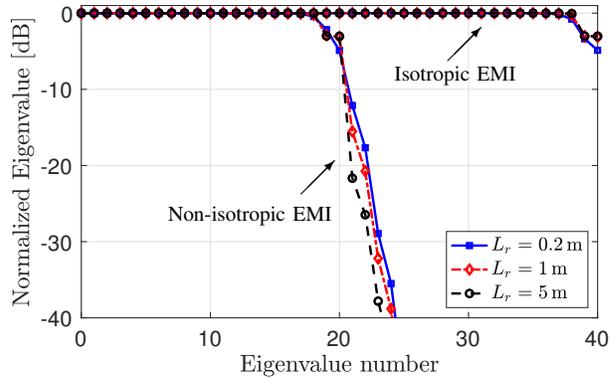

{To quantify the spatial correlation properties of noise samples $\{z_n^{(\rm{emi})}\}$, Fig.~\ref{eigenvalues_noise} plots the normalized eigenvalues in dB of ${\bf R}$ when $L_s = 0.2$\,m. We assume $\lambda=0.01$\,m, which yields $N_{\max} = 41$. We consider both isotropic and non-isotropic propagation conditions. In this latter case, we assume that the electromagnetic waves are uniformly distributed over the angular interval $\theta_r\in [\pi/3,2\pi/3 )$ and $\varphi_r\in [ -\pi,\pi)$. This is characterized by $f(\theta_r,\varphi_r) = \frac{\sin \theta_r}{2 \pi}$ from which it follows that $\rho(z) = \sinc \left(\frac{{z}}{\lambda}\right)$. In line with the observations made above, the figure shows that the eigenvalues of ${\bf R}$ in the isotropic case are similar but not \emph{exactly} the same. This means that the EMI samples exhibits some correlation, which diminishes as $L_r$ increases. This is not the case under non-isotropic conditions where large variations are observed for all the considered values of $L_r$. This shows that the EMI samples may be highly correlated when isotropic conditions are not verified.}


\section{Spectral Efficiency}\label{sec:spectral_efficiency}

%
Unlike OFDM, WDM guarantees the orthogonality condition among communication modes only when $L_r \to \infty$. This implies that, for finite values of $L_r$, {digital signal processing in the wavenumber domain is needed to deal with the interference. Next, we consider different solutions, and make comparisons in terms of SE.} In doing so, we impose the following constraint
\begin{align}\label{eq:power_constraint}
\frac{1}{L_s}\int_{-\frac{L_s}{2}}^{\frac{L_s}{2}} |i(s_z)|^2 ds_z \le P_s
\end{align}
and define the system SNR as 
\begin{align}\label{eq:snr-definition}
\mathrm{snr} = \frac{1}{\sigma^2_{{\rm emi}}}\frac{1}{L_s}\sum_{n=1}^N|x_n|^2 \mathop{=}^{(a)} \frac{(\kappa Z_0)^2}{\sigma^2_{{\rm emi}}}\frac{1}{L_s}\sum_{n=1}^N|\xi_n|^2\mathop{\le}^{(b)}\frac{P}{\sigma^2_{{\rm emi}}}
\end{align}
where $(a)$ follows from~\eqref{eq:signal_samples} and $(b)$ from $\sum_{n=1}^N|\xi_n|^2 /L_s\le P_s$, as obtained by plugging~\eqref{eq:current_signal} into~\eqref{eq:power_constraint}, and defining $P = (\kappa Z_0)^2 P_s$. Notice that $P_s$ is measured in A$^2$ and $P$ in [V$^2$/m$^2$]. {We assume that the EMI is isotropic, i.e., characterized by~\eqref{rho_r} or, equivalently, by~\eqref{eq:S_x}.}

\subsection{With Singular-Value Decomposition}
As any MIMO system, the {maximum SE} of WDM is given by~\eqref{eq:MIMO-capacity} and it can be achieved via the transceiver architecture in Fig.~\ref{figure_optimal_transceiver}. This requires the SVD of ${\widetilde{\bf H}} = {\bf L}^{-1}{\bf H}$ where ${\bf H}$ has entries given by~\eqref{eq:H_ni-free-space_linear} or~\eqref{eq:H_ni_frequency_domain_x}, and ${\bf L}$ is obtained from the Cholesky decomposition of the covariance matrix of 
${{\bf z}}\sim \mathcal{N}_{\mathbb{C}} ({\bf 0}_N, {\bf C})$ where {${\bf C} = \sigma^2_{{\rm emi}}{\bf R} + {\sigma^2_{{\rm hdw}}}{\bf I}_N$} and the elements of ${\bf R}$ are given by~\eqref{eq:correlation_matrices_entries_noise}. Particularly, ${\bf H}$ is a deterministic matrix, whose computation requires knowledge of the Green's function and the Fourier basis functions $\{{\boldsymbol\phi}_m(\vect{s})\}$ and $\{{\boldsymbol \psi}_n(\vect{r})\}$ given by~\eqref{eq:B0} and~\eqref{eq:B0-rx}, respectively. They all depend on the system parameters, e.g., wavelength $\lambda$, distance $d$, sizes $L_s$ and $L_r$. {Hence, we can assume that ${\bf H}$ is perfectly known at both sides of the communication link.} The same is true for ${\bf C}$, which also requires knowledge of $\sigma^2_{{\rm emi}}$ and $\sigma^2_{{\rm hdw}}$. Notice also that ${\bf H}$ and ${\bf C}$ are square matrices of size $N$, i.e., the number of communication modes. Since $N\le N_{\max} = 2\floor*{ \frac{L_s}{\lambda}}+1$, the maximum number of complex operations required for pre-processing $\bf x$ and post-processing $\bf y$ (through $\widetilde{\bf V}$ and $\widetilde{\bf U}$, respectively) is $2N_{\max}^2$. If $\lambda =0.01$\,m, then $N_{\max} = 41$ and $N_{\max} = 101$ for $L_s = 0.2$\,m and $L_s = 0.5$\,m, respectively. It increases to $N_{\max} = 401$ and $N_{\max} = 1001$ if $\lambda =0.001$\,m. {Hence, the computational complexity may become high for sources of large size and/or high frequency transmissions. Simpler solutions are thus discussed next.}

%
%

%

\subsection{With Linear Processing at the Receiver only}
Simpler schemes can be obtained by assuming that no pre-processing is performed at the source, i.e., $\widetilde{\bf V} = {\bf I}_N$, while linear processing is still applied at the receiver with $\widetilde{\bf U} =[\widetilde{\bf u}_1^{\Htran},\ldots,\widetilde{\bf u}_N^{\Htran}]^{\Htran}$. In this case, $\widetilde {\bf y} = \widetilde{\bf U}{\bf L}^{-1}{\bf y}$, ${\bf x} = \widetilde{\bf x}$, and $\widetilde{{\bf z}} = \widetilde{\bf U}{\bf L}^{-1}{{\bf z}}$ has independent and identically distributed Gaussian entries with $ \widetilde z_n \sim \mathcal{N}(0,\sigma_n^2)$ and $\sigma_n^2 = ||\widetilde{\bf u}_n||^2$. In scalar form, we obtain
\begin{align}\label{eq:IO-discrete-interference}
\widetilde y_n  = \underbrace{{\widetilde{\bf u}_n^{\Htran}}\widetilde{\bf h}_n x_n}_{\text{desired signal}}\; + \;\underbrace{\sum_{m=1,m\ne n}^{N} {\widetilde{\bf u}_n^{\Htran}}\widetilde{\bf h}_m \, x_m}_{\text{interference}} \;+ \; \widetilde z_n
\end{align}
where $\widetilde{\bf H} =[\widetilde{\bf h}_1^{\Htran},\ldots,\widetilde{\bf h}_N^{\Htran}]^{\Htran}$.
Assume that $x_n\sim \mathcal {CN}(0,p_n)$ and that the decoding is performed by treating the interference as Gaussian noise. For any fixed powers $\{p_n;n=1,\ldots,N\}$, the sum SE (measured in bits per channel use) is $\mathsf{SE} = \sum\limits_{n=1}^{N}\log_2(1 + \mathsf{SINR}_n)$
where
\begin{align}\label{eq:SE-iterativeWf}
\mathsf{SINR}_n = \frac{|\widetilde{\bf u}_n^{\Htran}\widetilde{\bf h}_n|^2 p_n}{\sum\limits_{m=1,m\ne n}^{N}|\widetilde{\bf u}_n^{\Htran}\widetilde{\bf h}_m|^2 p_m+ \widetilde{\bf u}_n^{\Htran}\widetilde{\bf u}_n}
\end{align}
is the signal-to-interference-plus-noise ratio (SINR). This is a generalized Rayleigh quotient with respect to $\widetilde{\bf u}_n$ and thus is maximized by the MMSE combiner, i.e., 
\begin{align}\label{eq:MMSE}
\widetilde{\bf u}_n^{\rm MMSE} = \left(\sum_{m=1}^Np_m\widetilde{\bf h}_m\widetilde{\bf h}_m^{\Htran}+\widetilde{\bf I}_N\right)^{-1}\!\!\!\!\widetilde{\bf h}_n.
\end{align}
{Despite optimal, $\widetilde{\bf u}_n^{\rm MMSE}$ requires a matrix inversion of size $N$, which may be too cumbersome when $N$ is large (e.g., high frequencies). To avoid the inverse computation, we can resort to the {maximum ratio (MR) combiner} given by $\widetilde{\bf u}_n^{\rm MR}=\widetilde{\bf h}_n$, which maximizes the power of the desired signal but neglects the interference. {A further simplification follows from Corollary~\ref{corollary1}, which shows that the communication modes become orthogonal as $L_r$ grows to infinity. Hence, the asymptotic optimal scheme for WDM is a bank of one-tap complex-valued multipliers, i.e., 
\begin{align}\label{eq:bank_of_multipliers}
[\widetilde{\bf u}_n]_m  =    \left\{{\begin{array}{cc}
    \widetilde{H}_{nn},& n=m \\
   0& n \ne m \\
  \end{array} } \right.
\end{align}
In the sequel, when WDM is used with~\eqref{eq:bank_of_multipliers}, we simply refer to it as WDM. Notice that it represents the simplest implementation of WDM.}}

The powers that maximize~\eqref{eq:SE-iterativeWf} for any choice of $\{\widetilde{\bf u}_n;n=1,\ldots,N\}$ can be obtained by means of the iterative waterfilling algorithm, whose convergence is not always guaranteed as it depends on the amount of interference (e.g.,~\cite{Yu2004}). To overcome this problem, we exploit Corollary~\ref{corollary1} and assume that $L_r$ is sufficiently large such that the interference can reasonably be neglected. In this case, the optimal powers can be computed as 
\begin{align}\label{eq:waterfilling_1}
{p_n^\star = \max\left(0,\mu - \frac{1}{|\widetilde{H}_{nn}|^2}\right)}
\end{align}
with $\sum_{n=1}^N p_n^\star  = P$. Unlike~\eqref{eq:water-filling}, the powers are computed with respect to $\{|\widetilde{H}_{nn}|^2; n=1,\ldots,N\}$.

\begin{figure}
\centering
\begin{subfigure}{.5\textwidth}
  \centering
	\begin{overpic}[width=1.\columnwidth,tics=10]{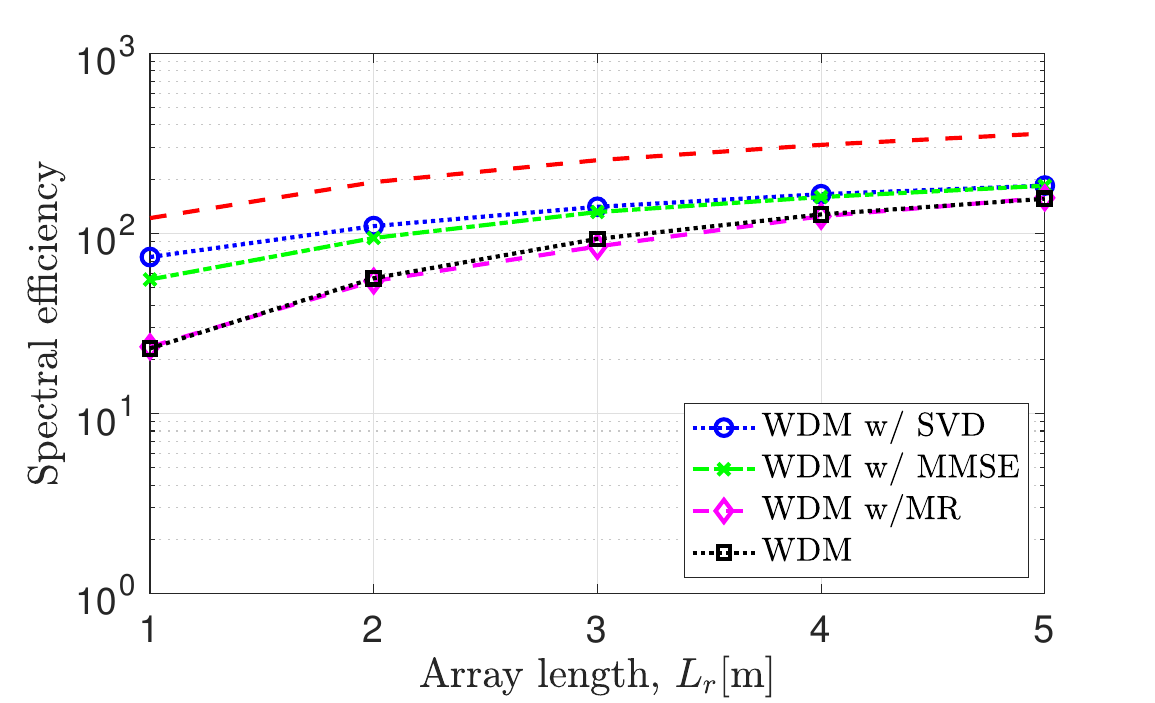}
	\put(28,51){\footnotesize Optimal scheme}
	\put(32,50){\vector(-1, -1){5}}
\end{overpic} 
                \caption{Distance $d = 5$\,m.}               \label{fig:spectralefficiency_rz5} 
\end{subfigure}%
\hspace{0.2cm}
\begin{subfigure}{.5\textwidth}
  \centering
	\begin{overpic}[width=1.\columnwidth,tics=10]{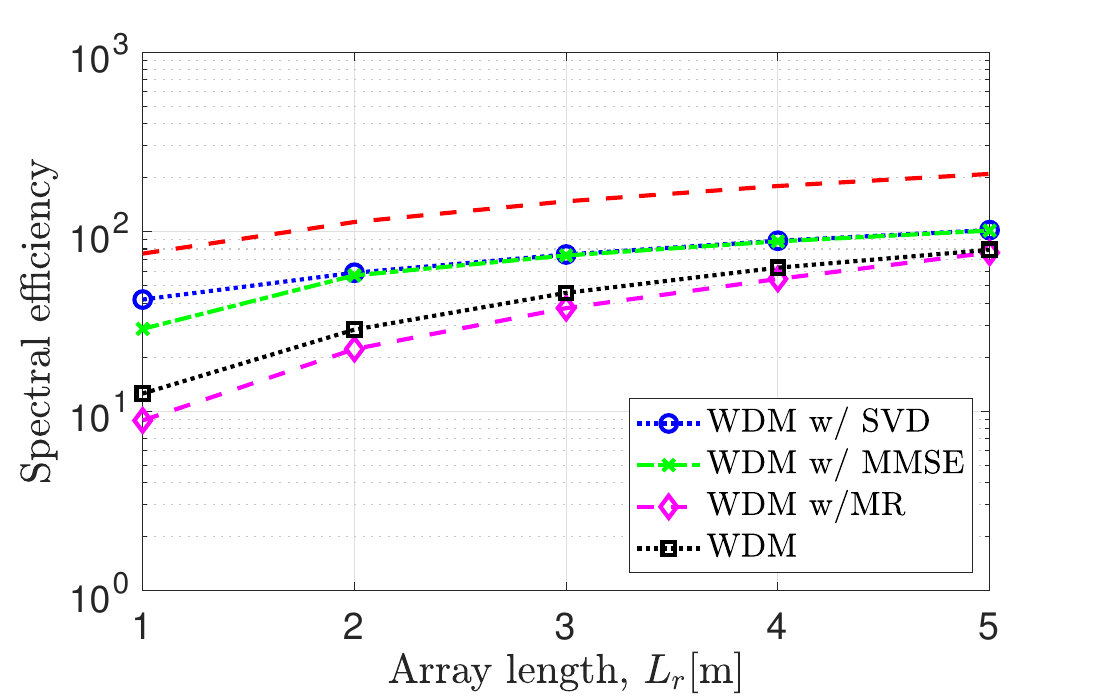}	
	\put(28,49){\footnotesize Optimal scheme}
	\put(32,48){\vector(-1, -1){5}}
\end{overpic}  
                \caption{Distance $d = 10$\,m.}
                \label{fig:spectralefficiency_rz10}
\end{subfigure}
        \caption{Sum SE (in bits per channel use) of WDM with different signal processing schemes. The optimal scheme is obtained through~\eqref{eq:86} --~\eqref{eq:88}, and is reported as a reference.}
        \label{fig:spectralefficiency_rz}
\end{figure}

\subsection{Numerical analysis}
Numerical results are used to quantify the sum SE of WDM with the different signal processing schemes of Sections V-A and V-B. We assume $L_s = 0.2$\,m and $\lambda = 0.01$\,m. From~\eqref{eq:max_number_of_modes}, the maximum number of communication modes is $N_{\max} = 41$ with spacing in the wavenumber domain of $2\pi/L_s = 31.41$\,rad/m. {We assume that all communication modes can be used for transmission.} Hence, we set $N = N_{\rm {\max}}$. The system SNR given by~\eqref{eq:snr-definition} is $\mathrm{snr} = 90$\,dB and the source power is $P_s = 10^{-7}$\,[A$^2$]. Accordingly, from Section~\ref{sec:electromagnetic_radiated_power} we have that the radiated power in~\eqref{UB_Prad} is $P_{\rm{rad}} \le 3.8\times 10^{-3}$ [W/m] and $\sigma_{\rm{emi}}^2 = 5.6\times 10^{-6}$ [V$^2$/m$^2$]. {We begin by neglecting the hardware noise and set $\sigma_{\rm{hdw}}^2=0$. Its impact will be taken into account later.} Comparisons are made with a communication system in which the basis sets $\{{\boldsymbol\phi}_m(\vect{s}); m=1,\ldots,N\}$ and $\{{\boldsymbol \psi}_n(\vect{r}); n=1,\ldots,N\}$ are chosen as the eigenfunctions of the channel operator~\eqref{eq:K_s}. As discussed in Section~\ref{sec:WDM}, this requires to select $\phi_n(s_z)$ in~\eqref{eq:B0} as the solution of the following eigenfunction problem (e.g., \cite{Miller:00}--~\cite{Wallace2008}):
\begin{align}\label{eq:86}
\gamma_n\phi_n(s_z) = \int_{-L_s/2}^{L_s/2} K_s(s_z, s_z^\prime) {\phi}_n(s_z^\prime)d s_z^\prime
\end{align}
with 
\begin{align}\label{eq:87}
K_s(s_z, s_z^\prime)= \int_{-L_r/2}^{L_r/2} g_z^*(r_z - s_z, d) g_z(r_z - s_z^\prime, d)d r_z.
\end{align}
The output basis function in~\eqref{eq:B0-rx} are then obtained as 
\begin{align}\label{eq:88}
 {\psi}_n(r_z) = \int_{-L_r/2}^{L_r/2}  g_z(r_z - s_z, d) \phi_n(s_z)d s_z.
 \end{align}
{With spatially uncorrelated noise, the above scheme is optimal and is reported as a reference.}

\begin{figure}[t!]
	\centering 
	\begin{overpic}[width=1\columnwidth,tics=10]{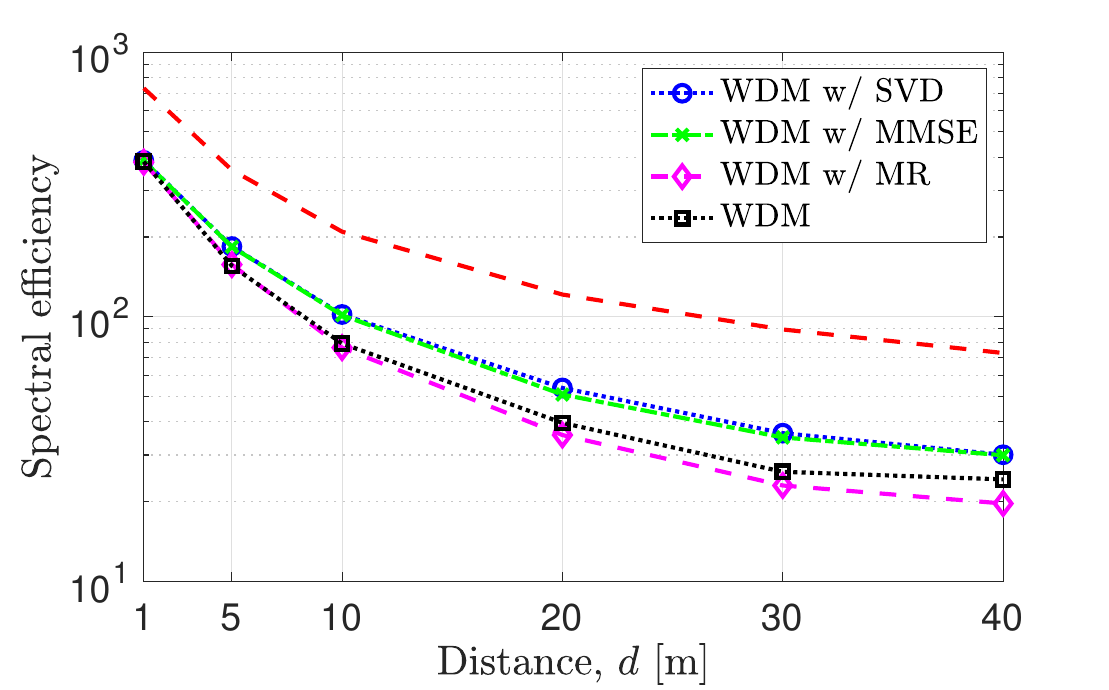}
	\put(32,46){\footnotesize Optimal scheme}
	\put(36,45){\vector(-1, -1){4}}
\end{overpic}
	\caption{Sum SE of WDM in bits per channel use as a function of distance when $L_r=5$\,m. }
	\label{fig:spectralefficiency_vs_distance}
\end{figure}

{Fig.~\ref{fig:spectralefficiency_rz} plots the SE as a function of $L_r$ when $d=5$ and $10$\,m. We see that WDM with SVD and MMSE processing achieves roughly the same SE for $L_r\ge 2$\,m. Compared to the optimal scheme, a substantial loss is observed. However, we recall that the superior performance of the optimal scheme is achieved at the cost of a prohibitively high implementation complexity. As expected, WDM with MR combining achieves the same SE as with SVD and MMSE only when $L_r$ grows and the interference reduces. As a rule of thumb, this happens approximately for $L_r \ge d$. We see that the simplest implementation of WDM, obtained  with~\eqref{eq:bank_of_multipliers}, provides the same SE with MR combining. Interestingly, it performs even better when $d=10$\,m. Fig.~\ref{fig:spectralefficiency_vs_distance} plots the SE as a function of the distance when $L_r=5$\,m. We see that SVD and MMSE provide the same performance irrespective of $d$. The loss incurred by using the simplest WDM increases as $d$ grows, due to the residual interference. The SE of all schemes reduces as $d$ increases since both the received power and the number of communication modes decrease. We notice that similar trends are observed if a uniform power allocation policy is used with WDM.
}

\subsection{Matched electromagnetic processing at the receiver}
\begin{figure}[t!]
	\centering 
	\begin{overpic}[width=1\columnwidth,tics=10]{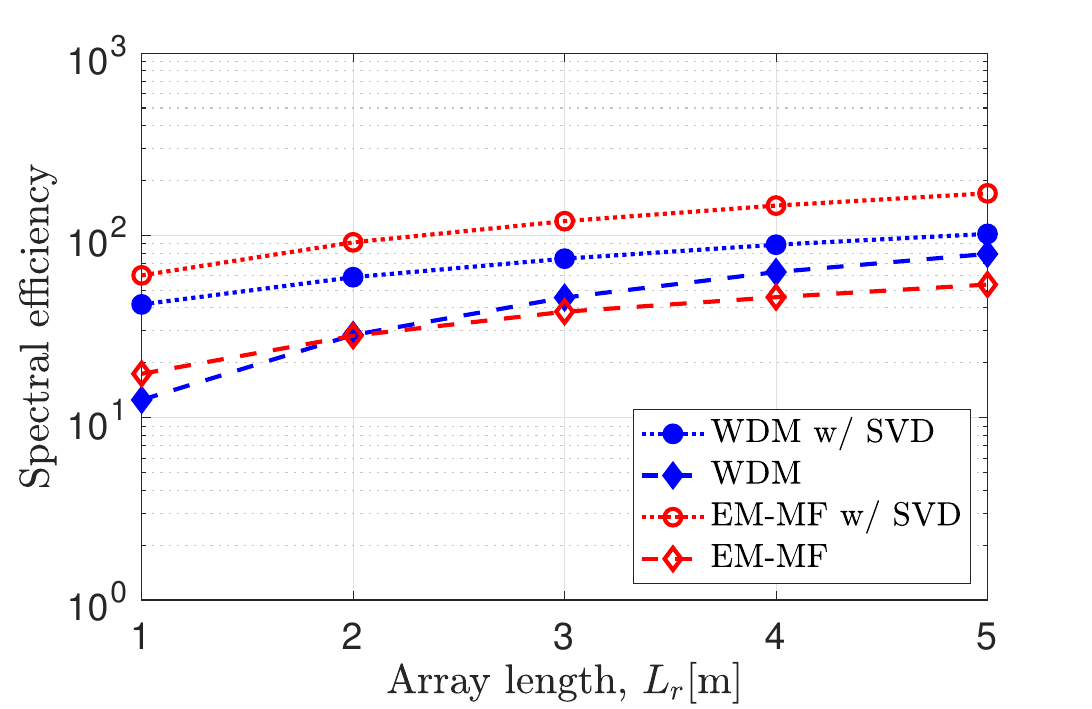}
\end{overpic}
	\caption{Comparison between WDM and EM-MF with different processing when $d=10$\,m. }
	\label{fig:spectralefficiency_vs_EM_MF}
\end{figure}

{Comparisons are now made with a communication scheme in which the Fourier basis functions $\{\phi_m(s_z); m=1,\ldots,N\}$ in~\eqref{eq:B0} are still used at the source but, at the receiver, the basis functions $\{\psi_n(r_z);n=1,\ldots,N\}$ are set equal to $\{\theta_m(r_z);n=1,\ldots,N\}$ in~\eqref{eq:optimal_RX_basis_functions}. This scheme is optimal in the sense that the receiver matches the effect of propagation channel over each $\phi_m(s_z)$, as it follows from~\eqref{eq:optimal_RX_basis_functions}. Hence, we call it electromagnetic matched-filter (EM-MF). Unlike the optimal scheme obtained through~\eqref{eq:86}--\eqref{eq:88}, EM-MF cannot remove the interference among communication modes. Hence, digital signal processing may be needed when $L_r$ is not sufficiently large. As for the optimal scheme, we stress that the implementation of EM-MF is challenging due to the functional dependence of the basis functions $\theta_m(r_z)$ on the Green's function. This is not the case with WDM. Fig.~\ref{fig:spectralefficiency_vs_EM_MF} reports the SE achieved by EM-MF when used with SVD or~\eqref{eq:bank_of_multipliers}. Compared to WDM, a substantial gain is observed with SVD whereas a similar SE is achieved with~\eqref{eq:bank_of_multipliers}. This shows that WDM is a valuable solution in terms of performance and implementation complexity.} 

\subsection{Comparisons with a fully-digital MIMO communication scheme}
{The developed framework is now used to make comparisons with a MIMO system in which only an integration of the electric field is performed over spatial segments of size $\delta$ that are equally spaced of $\Delta_s$ and $\Delta_r$ at the source and receiver, respectively. In this case, the functions $\{\phi_m(s_z); m=1,\ldots,N\}$ take the form $\phi_m(s_z) =  \frac{1}{\sqrt{\delta}}{\rm{rect}}\left(\frac{s_z- L_s/2 +(m - 1)\Delta_s}{\delta}\right)
$
where ${\rm{rect}}\left(\frac{s}{\delta}\right)$ is the rectangular function centered in zero and of duration $\delta$. At the receiver, we have that $\psi_n(r_z) =  {\rm{rect}}\left(\frac{r_z- L_r/2 +(n - 1)\Delta_r}{\delta}\right)$.
Hence, the output samples are
\begin{align}\notag
y_n &= \int_{-\frac{L_r}{2}}^{\frac{L_r}{2}}  \psi_n^*(r_z)  y(d,0,r_z) dr_z + z_n^{({\rm hdw})}\\&=\int_{-\frac{L_r}{2} - \frac{\delta}{2}+(n - 1)\Delta_r}^{-\frac{L_r}{2} + \frac{\delta}{2}+(n - 1)\Delta_r } y(d,0,r_z)dr_z + z_n^{({\rm hdw})}\label{eq:dft-rx-MIMO}
\end{align}
which is simply an integration over the space interval where the dipole $n$ of length $\delta$ is located. Notice that~\eqref{eq:H_ni} reduces to
\begin{align}\label{eq:H_ni-free-space_linear-MIMO}
\!\!H_{nm} =  \int_{-\frac{L_r}{2} - \frac{\delta}{2}+(n - 1)\Delta_r}^{-\frac{L_r}{2} + \frac{\delta}{2}+(n - 1)\Delta_r }\int_{-\frac{L_s}{2} - \frac{\delta}{2}+(m - 1)\Delta_s}^{-\frac{L_s}{2} + \frac{\delta}{2}+(m - 1)\Delta_s } \hspace{-1cm}g_z(r_z-s_z,d)   d s_z d r_z
\end{align}
while 
the matrix ${\bf R}$ in~\eqref{eq:correlation_matrices_entries_noise} has now entries
\begin{align}\label{eq:correlation_matrices_entries_noise_MIMO}
\!\!\left[ {\bf R}\right]_{nm} \!\!= \int_{-\frac{L_r}{2} - \frac{\delta}{2}+(n - 1)\Delta_r}^{-\frac{L_r}{2} + \frac{\delta}{2}+(n - 1)\Delta_r }\int_{-\frac{L_r}{2} - \frac{\delta}{2}+(m - 1)\Delta_r}^{-\frac{L_r}{2} + \frac{\delta}{2}+(m - 1)\Delta_r } \hspace{-1cm}\rho\left({r_z-r_z^\prime}\right) dr_zdr_z^\prime.\end{align}}
For the setup of Fig.~\ref{figure_linear_arrays}, the number of RF chains at the source (say $N_s$) and receiver (say $N_r$) is given by $N_s = \floor*{\frac{L_s}{\Delta_s}}+1$ and $N_r = \floor*{\frac{L_r}{\Delta_r}}+1$.
For $L_s = 0.2$\,m and $L_r = 5$\,m, if $\Delta_s=\Delta_r=\Delta = \lambda/2$ with $\lambda = 0.01$\,m, we have that $N_s = 41$ and $N_r = 201$. This is different from WDM where the RF chains required at both sides are given by~\eqref{eq:max_number_of_modes}.
If $L_s = 0.2$\,m and $\lambda = 0.01$, we obtain $N_{\max} = 41$ such that they are approximately reduced by a factor $5$. {We conclude by observing that the above model emulates a MIMO system in which the source and receiver are made of {elementary dipoles of size $\delta$ much smaller than the wavelength $\lambda$~\cite{Balanis-2012-antenna}},\footnote{{Such that the current distribution can be assumed to be uniform over the dipole length. In practice, dipoles of size in the order of $\delta \le \lambda/10$ should be considered for the model to be valid.}} equally spaced of $\Delta_s$ and $\Delta_r$. Each dipole is connected to an RF chain. Since no other processing is performed except for the the integration, we simply refer to it as `MIMO' in the sequel.} 

\begin{figure}
\centering
\begin{subfigure}{0.5\textwidth}
  \centering
	\begin{overpic}[width=1.\columnwidth,tics=10]{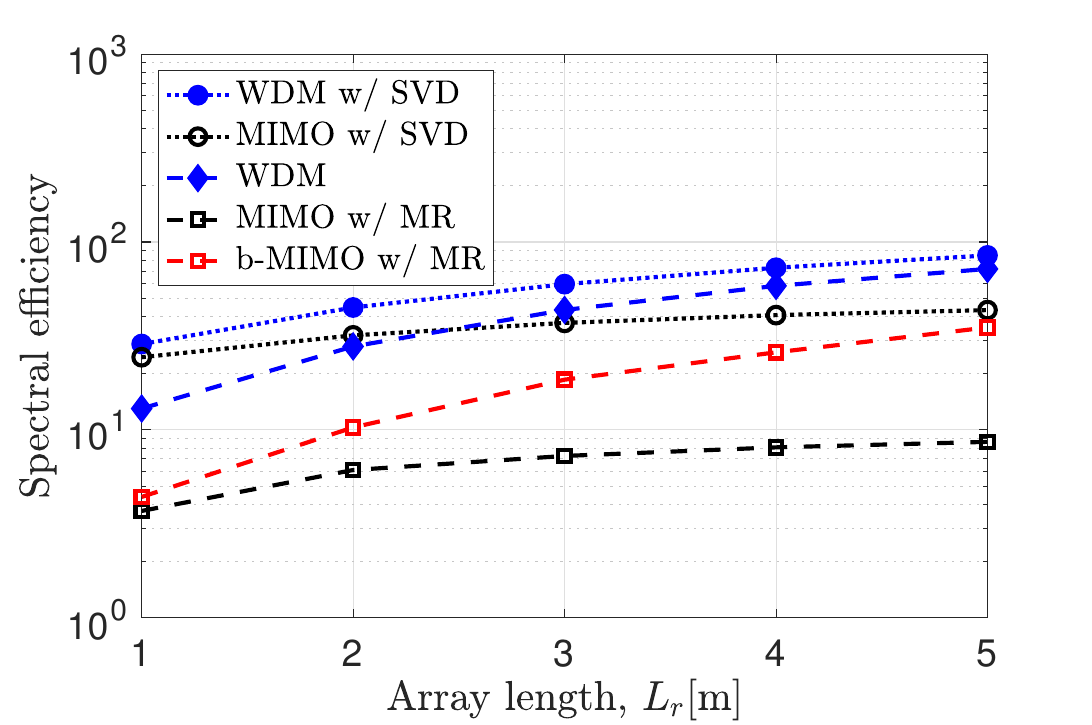}
\end{overpic} 
\caption{$N_s=N_r=N$, $d= 10$\,m.}  \vspace{0.2cm}
\label{fig:spectralefficiency_rzMIMO} 
\end{subfigure}
\begin{subfigure}{.5\textwidth}
  \centering
	\begin{overpic}[width=1.\columnwidth,tics=10]{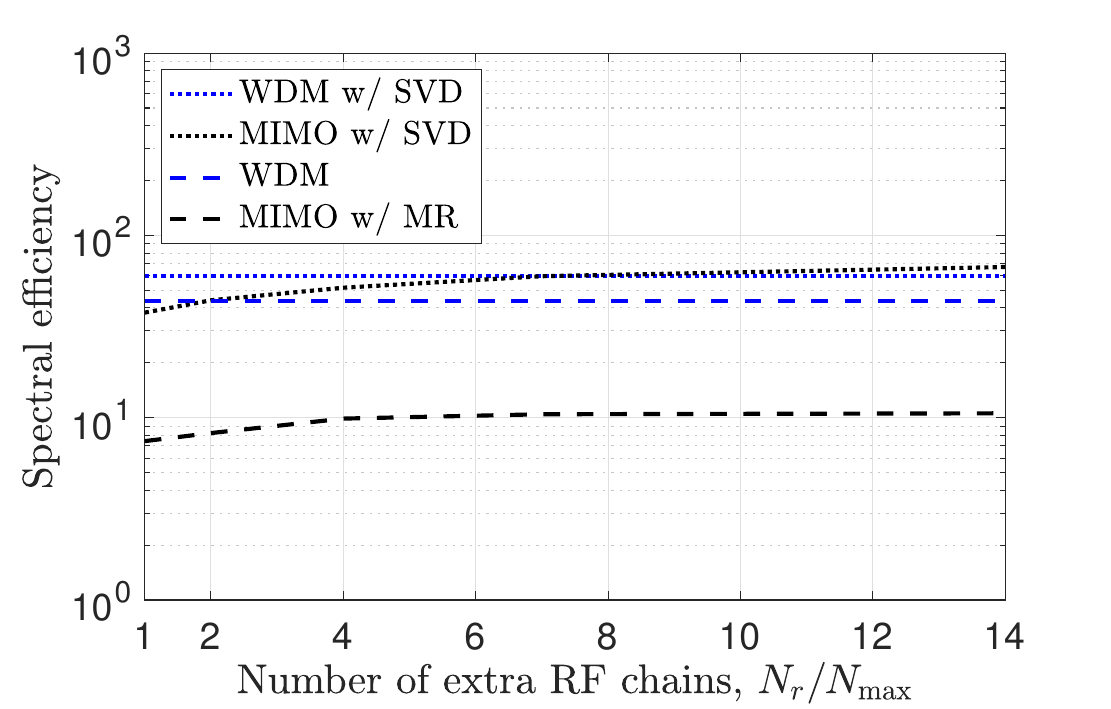}	
\end{overpic}  
\caption{$L_r = 3$\,m and $d= 10$\,m.}
\label{fig:spectralefficiency_vs_nrf}
\end{subfigure}
\caption{Comparisons with a MIMO system with the same or different number of RF chains of WDM. The dipole size in the MIMO system is $\delta = \lambda/10$.}
\end{figure}

Fig.~\ref{fig:spectralefficiency_rzMIMO} plots the SE when $d=10$\,m. To take into account the impact of the hardware noise introduced by the RF chains, we assume that $\sigma_{\rm{hdw}}^2 = 10\sigma_{\rm{emi}}^2$. We set\footnote{Notice that the performance of the MIMO system depends on the size $\delta$ of the rectangular function.} {$\delta = \lambda/10$} and assume $N_s = N_r = N_{\max} = 41$, i.e., the same number of RF chains is used with WDM and MIMO. This is achieved by fixing $\Delta_s = \lambda/2$ while setting $\Delta_r = \floor*{ \frac{\lambda}{2} \frac{L_r}{L_s}}$. The results from Fig.~\ref{fig:spectralefficiency_rzMIMO} show that {WDM with SVD provides the highest SE. Interestingly, the simplest implementation of WDM highly outperforms MIMO with MR combining for any value of $L_r$ and MIMO with SVD for $L_r >3$}. The gain is even larger if MIMO is used with~\eqref{eq:bank_of_multipliers} (not shown for space limitations). {Comparisons are also made with a beamspace MIMO (b-MIMO) scheme~\cite{Sayeed2013a} in which the transmitted symbols $\{x_m\}$ are pre-processed by an $N-$dimensional inverse discrete Fourier transform (IDFT) matrix and the output samples $\{y_n\}$ in~\eqref{eq:dft-rx-MIMO} are post-processed by an $N-$dimensional DFT matrix. We see that, with MR combining b-MIMO performs better than MIMO but WDM is still better. Notice that, with SVD processing, b-MIMO achieves the same SE (not reported for convenience) as MIMO since the DFT matrices are unitary.}

Fig.~\ref{fig:spectralefficiency_vs_nrf} plots the SE as a function of $N_r/N_{\max}$ (i.e., number of extra radio frequency chains used in the MIMO system) when $L_r = 3$\,m, $d= 10$\,m and $\sigma_{\rm{hdw}}^2 = 10\sigma_{\rm{emi}}^2$. {We see that a MIMO system achieves marginal gains compared to WDM with SVD processing only when $N_r/N > 8$, i.e., at the price of a much higher implementation complexity. On the contrary, MIMO with MR combining performs always much worse than WDM even if $N_r/N_{\max}$ grows.}

\section{Conclusions and Future Research Directions}\label{sec:conclusions}

Building on prior analyses (e.g., \cite{Miller:00,Marengo2008,Wallace2008}), we provided a MIMO representation of the electromagnetic wave propagation problem between two spatially-continuous volumes, 
which can be directly used by communication theorists as a baseline for developing and studying optimal or suboptimal (but implementable) holographic MIMO communication schemes. Particularly, we used it to develop a communication scheme for the system setup in Fig.~\ref{figure_linear_arrays}; that is, linear sources and receivers in LoS propagation conditions. Inspired by OFDM, we made use of Fourier basis functions and proposed a WDM communication scheme that operates in the wavenumber domain and makes use of Fourier transform operations directly at the electromagnetic level. Conventional tools of linear systems theory were used to understand the interplay among the different system parameters in terms of number of communication modes and level of interference. Unlike OFDM, the orthogonality among the communication modes (in the wavenumber domain) is achieved with WDM only when the size of the receiver becomes infinitely large, due to the unbounded support of the channel response in the spatial domain. Different communication architectures were thus used and designed to deal with the interference. Numerical results showed that, for the investigated scenarios, the optimal SVD architecture with water-filling and the linear MMSE receiver with suboptimal power allocation achieve the same SE, while {the simplest implementation performs sufficiently well starting from a receiver size of the same order of the distance from the source.}

Future research directions for the proposed WDM scheme are clearly represented by considering scenarios where some of the underlying assumptions of Fig.~\ref{figure_linear_arrays} are not valid. A first extension is to study the effects of multipath propagation and LoS blockage. Since in practice the source and receiver are not parallel, the impact of arbitrary positions and orientations should be investigated. {This would open the door for multi-user communications in which multiple sources (arbitrarily located in the area) transmit to a fixed receiver; the different location and orientation of sources may potentially be exploited to remove the interference among multiple data streams by means of linear or non-linear processing.} A natural extension is also the design of the WDM scheme when two-dimensional (rather one-dimensional) surfaces are used for transmission and reception~\cite{Dardari-2020}; this provides additional flexibility in terms of generating the current densities and processing the impinging electric fields. {Finally, we notice that, although the idea of holographic communications has its roots in LoS propagation conditions at high frequencies, the concept can be ``scaled'' accordingly to lower frequencies where non-LoS propagation conditions (rich scattering environment) would be a must to compensate for the lower number of communication modes.}


\section*{Appendix A}\label{AppendixA}
%

From~\cite[Ch. 14]{OrfanidisBook}, the radiated power $\mathsf{P_{rad}}$ is given by $\mathsf{P_{rad}} = \lim_{r \to \infty} \int_{\Omega} P_r  r^2 d \Omega$
where \begin{align}\label{eq:Pr}
P_r = \frac{1}{2Z_0} {\bf e}^{\Htran}({\bf{r}}){\bf e}({\bf{r}})
\end{align}
is the radial component of the Poynting vector \cite[Ch. 14]{OrfanidisBook}, ${\bf e}({\bf{r}})$ is the \textit{radiation} electric field with $r = ||{\bf r}||$, and $\Omega$ is the solid angle of $4 \pi$ steradians. When $r \to \infty$, ${\bf e}({\bf{r}})$ reduces to
\begin{equation}
\label{FieldCurrent_FFA}
{\bf e}(\vect{r})=\imagunit \kappa Z_0 \frac{e^{ \imagunit \kappa r}}{4 \pi r}  \left(\mathbf{I}-\widehat{\bf r} \widehat{\bf r}^{\Htran}\right) \int_{V_s} {\bf j}(\vect{s}) e^{ -\imagunit {\boldsymbol \kappa}^{\Ttran}(\theta_s, \varphi_s) {\bf s}}\, d\vect{s}
\end{equation}
since the Green's function \eqref{DyadicGF.2} can be approximated as 
\begin{equation}
\label{DGF_FFA}
\vect g(\mathbf{r},\mathbf{s}) \approx \frac{1}{4 \pi}\frac{e^{ \imagunit \kappa r}}{ r}  \left(\mathbf{I}-\widehat{\bf r} \widehat{\bf r}^{\Htran}\right)e^{ -\imagunit {\boldsymbol \kappa}^{\Ttran}(\theta_s, \varphi_s) {\bf s}}.
\end{equation}
Plugging~\eqref{FieldCurrent_FFA} into~\eqref{eq:Pr} yields
\begin{align}\notag
P_r&=\dfrac{Z_0}{8 \lambda^2} \dfrac{1}{r^2}\left\| \left(\mathbf{I}-\widehat{\bf r} \widehat{\bf r}^{\Htran}\right) \int_{V_s} {\bf j}(\vect{s}) e^{ -\imagunit {\boldsymbol \kappa}^{\Ttran}(\theta_s, \varphi_s) {\bf s}}\, d\vect{s} \right\|^2\\&
\le \dfrac{Z_0}{8 \lambda^2} \dfrac{1}{r^2}\left\| \int_{V_s} {\bf j}(\vect{s}) e^{ -\imagunit {\boldsymbol \kappa}^{\Ttran}(\theta_s, \varphi_s) {\bf s}}\, d\vect{s} \right\|^2.\label{PoyntingRadial}
\end{align}
From \eqref{PoyntingRadial}, we thus have that $\mathsf{P_{rad}}$ can be upper bounded as
\begin{align}\notag
\mathsf{P_{rad}}& \le \dfrac{Z_0}{8 \lambda^2} \int_{\Omega} \left\| \int_{V_s} {\bf j}(\vect{s}) e^{ -\imagunit {\boldsymbol \kappa}^{\Ttran}(\theta_s, \varphi_s) {\bf s}} \, d\vect{s} \right\|^2 d \Omega\\&= 4\pi\dfrac{Z_0}{8 \lambda^2}  \iint_{V_s} {\bf j}^{\Htran}(\vect{s}_1){\bf j}(\vect{s}_2) \rho(\vect{s}_1-\vect{s}_2)    d\vect{s}_1 d\vect{s}_2\label{UB1}
\end{align}
where $\rho(\cdot)$ is given in~\eqref{rho_r}.
By applying the Cauchy-Schwarz inequality, we obtain~\eqref{UB_Prad}.

\section*{Appendix B}
{We start observing that the electric field generated by $\phi_{k}(s_z)$ along the $z-$axis is given by
\begin{equation}
e_{z}(d,0,r_z;m)  = \imagunit \kappa Z_0 I_m \int_{-\frac{L_s}{2}}^{\frac{L_s}{2}} g_z(r_z-s_z,d)  \phi_{m}(s_z) d s_z.
\end{equation}
In explicit form, the electric field reads
\begin{equation}
e_{z}(d,0,r_z;m)  = \frac{\imagunit \kappa Z_0 I_m}{\sqrt{L_s}} \int_{-\frac{L_s}{2}}^{\frac{L_s}{2}}  \frac{d^2}{4 \pi} \frac{e^{\imagunit \kappa \sqrt{(r_z-s_z)^2 + d^2}}}{\left((r_z-s_z)^2 + d^2\right)^{3/2}} e^{\imagunit\frac{2\pi} {L_s}\big(m -1 - {\frac{N- 1}{2}}\big)s_z} d s_z. 
\end{equation}
Notice that when $L_s \ll d$ it can be approximated as
\begin{align}\notag
e_{z}(d,0,r_z;m)&\approx \imagunit \kappa Z_0 I_m \sqrt{L_s}\dfrac{d^2}{4 \pi } \dfrac{ e^{j\kappa\sqrt{d^2+r_z^2}}}{(d^2+r_z^2)^{3/2}} \\ &\mathrm{sinc}\left(\dfrac{L_s}{\lambda} \dfrac{r_z}{\sqrt{d^2+r_z^2}} -\bigg(m -1 - {\frac{N- 1}{2}}\bigg)\right).\label{eq:electric_field_m}
\end{align}
Since $\cos (\theta) = \frac{r_z}{\sqrt{d^2+r_z^2}}$ and $\sin(\theta) = \frac{d}{\sqrt{d^2+r_z^2}}$,
we may rewrite 
\begin{equation}
e_{z}(d,0,r_z;m)\approx \imagunit \frac{\kappa Z_0 I_m \sqrt{L_s}}{4 \pi d}e^{j\kappa\sqrt{d^2+r_z^2}}\zeta(\theta;m)
\end{equation}
where 
\begin{equation}\label{eq:E.9}
\zeta(\theta;m) = \sin^3(\theta)\mathrm{sinc}\left(\dfrac{L_s}{\lambda} \cos(\theta) -\bigg(m -1 - {\frac{N- 1}{2}}\bigg)\right).
\end{equation}
The normalized radiation pattern is given by
\begin{equation}\label{eq:E.10}
 \zeta^2(\theta_k;m) = \left(\frac{4 \pi d_k}{\kappa Z_0 I_m \sqrt{L_s}}\right)^2\left |e_{z}(d_k,0,r_z;m)\right |^2.
\end{equation}
}
\section*{Appendix C}
Define $\frac{\kappa_z^\prime}{2\pi} = \frac{\kappa_z}{2\pi}-\frac{m - 1 - (N-1)/2}{L_s}$ and rewrite the integral in~\eqref{eq:H_ni_frequency_domain_x} as follows
\begin{equation}\label{eq:B20}
\int_{-\infty}^\infty \!\!\!\!G_z\left( {\kappa_z^\prime} + 2\pi\frac{m - 1 - (N-1)/2}{L_s},d\right) Q_{n-m}(\kappa_z^\prime)  d\kappa_z^\prime
\end{equation}
with $Q_{i}(\kappa_z^\prime)= {\rm sinc}\left(\frac{\kappa_z^\prime}{2\pi} L_s\right) \sinc\left(\left(\frac{\kappa_z^\prime}{2\pi}+\frac{i}{L_s}\right) L_r\right).$
By using Parseval,~\eqref{eq:B20} becomes 
\begin{align}\label{eq:B39}
 \int_{-\infty}^\infty \!\!\!\!g_z\left(z^\prime,d\right) e^{\imagunit 2\pi\frac{m - 1 - (N-1)/2}{L_s}z^\prime} q_{n-m}(z^\prime)  dz^\prime
\end{align}
with $q_{i}(z^\prime) =  \frac{(2\pi)^2}{{L_sL_r}}{\rm rect}\Big(\frac{z^\prime}{L_s}\Big) \otimes {\rm rect}\Big(\frac{z^\prime}{L_r}\Big)e^{-\imagunit \frac{2\pi}{L_s}iz^\prime}$.
Assume now that the conditions in~\eqref{eq:B31} and~\eqref{eq:B30} are satisfied. In these circumstances, the integral in~\eqref{eq:B39} reduces to
\begin{align}\label{eq:B4.10}
\int_{-z_g}^{z_g} \!\!\!\!g_z\left(z^\prime,d\right) e^{\imagunit 2\pi\frac{m - 1 - (N-1)/2}{L_s}z^\prime} q_{n-m}(z^\prime)  dz^\prime.
\end{align}
For $|z^\prime| \le z_g$, it is easy to show that $q_{i}(z^\prime) = \frac{(2\pi)^2}{{L_r}}$ for $i=0$ and $q_{i}(z^\prime)=0$ for $i\ne 0$. Plugging this result into~\eqref{eq:B4.10} produces
\begin{align}\label{eq:B45}
\left\{{\begin{array}{cc}
  \frac{(2\pi)^2}{{L_r}} \int_{-z_g}^{z_g} g_z\left(z^\prime,d\right) e^{\imagunit 2\pi\frac{n - 1 - (N-1)/2}{L_s}x^\prime} dz^\prime  & n=m \\
   0& n \ne m.\\
  \end{array} } \right.
\end{align}
By noticing that, under the condition in~\eqref{eq:B30},
\begin{align}\label{eq:B60}
\int_{-z_g}^{z_g} \!\!g_z\left(z^\prime,d\right) e^{\imagunit 2\pi\frac{n - 1 - (N-1)/2}{L_s}z^\prime} dz^\prime = G_z\left(\frac{2\pi}{L_s}\left(n - N/2 -1\right) , d\right)
\end{align}
the orthogonality condition in~\eqref{eq:orthogonality-condition} easily follows.

\bibliographystyle{IEEEtran}
\bibliography{IEEEabrv,refs}

\end{document}